\documentclass[useAMS,usenatbib]{mn2e}
\pdfoutput=1
\usepackage{graphicx}

\begin{document}

\title[SMC star clusters]{The VMC survey - XV. The Small Magellanic
  Cloud--Bridge connection history as traced by their star cluster
  populations\thanks{Based on observations obtained with VISTA at ESO
    under programme ID 179.B-2003.}}

\author[A. E. Piatti et al.]{Andr\'es E. Piatti$^{1,2}$\thanks{E-mail: 
andres@oac.uncor.edu}, Richard de Grijs$^{3,4}$, Stefano Rubele$^{5}$, Maria-Rosa L. Cioni$^{6,7,8}$,
\newauthor Vincenzo Ripepi$^{9}$ and Leandro Kerber$^{10}$ \\
$^1$Observatorio Astron\'omico, Universidad Nacional de C\'ordoba, Laprida 854, 5000, 
C\'ordoba, Argentina\\
$^2$Consejo Nacional de Investigaciones Cient\'{\i}ficas y T\'ecnicas, Av. Rivadavia 1917, C1033AAJ,
Buenos Aires, Argentina \\
$^3$Kavli Institute for Astronomy and Astrophysics, Peking University, Yi He Yuan Lu 5, Hai Dian District, Beijing 100871, China\\
$^4$Department of Astronomy, Peking University, Yi He Yuan Lu 5, Hai Dian District, Beijing 100871, China\\
$^5$INAF, Osservatorio Astronomico di Padova, vicolo dell'Osservatorio 5, I-35122 Padova, Italy\\
$^6$Universit\"{a}t Potsdam, Institut f\"{u}r Physik und Astronomie, Karl-Liebknecht-Str. 24/25, 14476 Potsdam, Germany\\
$^7$Leibniz-Institut f\"{u}r Astrophysik Potsdam, An der Sternwarte 16, 14482 Potsdam, Germany\\
$^8$University of Hertfordshire, Physics, Astronomy and Mathematics, College Lane, Hatfield AL10 9AB, 
United Kingdom\\
$^9$INAF-Osservatorio Astronomico di Capodimonte, Via Moiariello 16, 80131, Naples, Italy\\
$^{10}$Universidade Estadual de Santa Cruz, Rodovia Ilh\'eus--Itabuna, km 16, 45662-200 Ilh\'eus, Bahia, Brazil\\
}

\maketitle

\begin{abstract} 
We present results based on $YJK_{\rm s}$ photometry of star clusters
located in the outermost, eastern region of the Small Magellanic Cloud
(SMC). We analysed a total of 51 catalogued clusters whose
colour--magnitude diagrams (CMDs), having been cleaned from field-star
contamination, were used to assess the clusters' reality and estimate
ages of the genuine systems. Based on CMD analysis, 15 catalogued 
clusters
were found to be possible non-genuine
aggregates. We investigated the properties
 of 80\% of the catalogued
clusters in this part of the SMC by enlarging our sample with
previously obtained cluster ages, adopting a homogeneous scale for
all. Their spatial distribution suggests that the oldest clusters,
log($t$ yr$^{-1}$) $\ge$ 9.6, are in general located at greater
distances to the galaxy's centre than their younger counterparts --
9.0 $\le$ log($t$ yr$^{-1}$) $\le$ 9.4 -- while two excesses of clusters
are seen at log($t$ yr$^{-1}$) $\sim$  9.2 and log($t$ yr$^{-1}$)
$\sim$ 9.7. We found a trail of younger clusters which follow the
Wing/Bridge components. This long spatial sequence does not only
harbour very young clusters, log($t$ yr$^{-1}$) $\sim$ 7.3, but it
also hosts some of intermediate ages, log($t$ yr$^{-1}$) $\sim$
9.1. The derived cluster and field-star formation frequencies as a function of age 
are different. The
most surprising feature is an observed excess of clusters with ages of
log($t$ yr$^{-1}$) $<$ 9.0, which could have been induced by
interactions with the LMC.
\end{abstract}

\begin{keywords}
techniques: photometric -- galaxies: individual: SMC -- Magellanic
Clouds
\end{keywords}

\section{Introduction}

The near-infrared VISTA Survey of the Magellanic Clouds (VMC) system
\citep{cetal11} has been designed to obtain three epochs of data in
the near-infrared passbands $Y$ and $J$, and 12 epochs in the
 $K_{\rm s}$, in order to reach a nominal survey
depth of $Y$ = 21.9, $J$ = 21.4 and $K_{\rm s}$ = 20.3 mag
(at a signal-to-noise ratio of 10) for individual tiles of 
$\sim1.5$ deg$^2$ in size. The VMC survey covers a
total area of $\sim170$ deg$^2$, comprising the Large Magellanic Cloud
(LMC), the Small Magellanic Cloud (SMC), the Magellanic Bridge and a
few tiles covering the Magellanic Stream. Individual epochs have
exposure times of 800 s ($Y$ and $J$) and 750 s ($K_\mathrm{s}$). The
multi-epoch observations allow us to minimize variability effects in
colours, particularly for bright objects, and guarantee homogeneous
observing conditions among different passbands for a given
epoch. Seeing constraints, imposed for the purpose of homogenizing
crowded and uncrowded field observations, range between 1.0$\arcsec$
and 1.3$\arcsec$ ($Y$), 0.9$\arcsec$ and 1.2$\arcsec$ ($J$), and
0.8$\arcsec$ and 1.0$\arcsec$ ($K_s$) and may exceed those values by 
10\%  according to observing policy. The averaged $K_{\rm s}$
magnitudes resulting from the multi-epoch observations for RR Lyrae
and Cepheid variable stars, with an accuracy of a few hundredths of a
magnitude, will allow us to unveil the 3D structure of the Magellanic
system \citep{retal12b,metal14c}. On the other hand, stacking of these 
observations allows us to
detect stars at the oldest main-sequence turn-off (MSTO) and derive
the star-formation history (SFH) across the system with unprecedented
quality \citep{retal12}.

VMC data also allow us to perform a thorough study of the Magellanic
Cloud (MC) star cluster population based on a homogeneous
determination of star cluster parameters \citep{petal14b}. In
addition, we can firmly establish whether the field star population
has experienced the same or a similar SFH as the star cluster system.
The addition of near-infrared photometry to existing optical
photometry is aimed at disentangling the scatter that currently exists
in the age--metallicity distribution \citep[e.g.][]{p11b,pg13}, also
enabling us to constrain extinction variations more precisely.
Wide-field VMC data will produce a complete census of the cluster
population that will be used to draw statistically robust conclusions
about  their differences between the Clouds. In addition, the search
for new clusters based on stellar surface density distributions will
work better in the Clouds than in the Galaxy because of the lower
confusion along the line of sight.

The outermost, eastern region of the SMC and the associated onset of
the Bridge are of significant scientific interest in the context of a
putative scenario involving interactions between the MCs, which may be
traced spatially and temporally. Indeed, the purpose of this paper is
to present a photometric analysis of the catalogued clusters located
in that region based on the most complete VMC data set to date. The
available photometric data allow us to confirm the physical reality of
the catalogued star clusters and (for the genuine clusters) estimate
their fundamental parameters. We also investigate the cluster
frequency in the region of interest to assess whether star clusters
and field stars have evolved as a coupled system.

This paper is organised as follows. The VMC data collection and
reduction is presented in Section 2, while the cluster sample is
described in Section 3. Sections 4 and 5 deal with the cleaning
procedure pertaining to the clusters' colour--magnitude diagrams and
the astrophysical property estimates for the clusters,
respectively. In Section 6 we discuss the photometric results, derive
the cluster frequency and compare it to the field star formation
history. We summarize our conclusions in Section 7.

\section{VMC data collection and reduction}

The VMC survey \citep{cetal11} has collected more than half of the
observations scheduled as part of a global observing campaign with the
VISTA telescope, where observations for five other ESO public surveys
are also carried out in service mode \citep{aetal13}. The survey's SMC
and Magellanic Bridge areas will eventually cover 42 deg$^2$ and 21
deg$^2$ (27 and 13 VISTA tiles), respectively. One tile covers
uniformly an area of $\sim$ 1.5 deg$^2$, representing a mosaic of six
paw-print images in a given passband ($YJK_{\rm s}$). Figure 1 shows
the distribution of the VMC tiles superimposed on that of the star
clusters (dots) catalogued by \citet[][hereinafter B08]{betal08} in
the SMC and the western Bridge regions. On the other hand, viewing the
SMC as a triaxial galaxy (and adopting the declination, Dec, right
ascension, RA, and line-of-sight as the three axes), \citet{cetal01}
found axial ratios of approximately 1:2:4. Based on this result, and
to describe the clusters' spatial distribution, \citet{petal07d} used
an elliptical rather than a spherical framework so as to reflect more
meaningfully the flattening of the galaxy. This reference system --
with one of its axes parallel to the SMC Bar and the other
perpendicular to that direction -- seems more appropriate to describe
the clusters' age and metallicity distributions than one with axes
along the right ascension and declination directions \citep[see,
  e.g.,][]{pietal08,p11b,p12a}. Two ellipses with semimajor axes of
2.4$\degr$ and 6.0$\degr$ are also shown. The ellipses are centred at
RA = 00h 52m 45s, Dec = $-$72$\degr$ 49$\arcmin$ 43$\arcsec$ (J2000)
\citep{cetal01} and have axis ratios, $b/a = 1/2$. Thus
the clusters' spatial distribution correlates better with a
pseudo-elliptical (projected) distance coordinate measured from the
galactic centre than with the radial distance, or with distances
defined along the right ascension or declination axes.

The VMC data are processed with the VISTA Data Flow System pipeline,
version 1.3 \citep[VDFS][]{ietal04}, and calibrated to match the VISTA
photometric system, which is close to the Vegamag system. We extracted
the observational data from the VISTA Science Archive
\citep[VSA,][]{cetal12}. The processed paw-print images were used to
derive the effective point-spread functions (PSFs) using the {\sc
  iraf}\footnote{IRAF is distributed by the National Optical Astronomy
  Observatories, which is operated by the Association of Universities
  for Research in Astronomy, Inc., under contract with the
  U.S. National Science Foundation.}/{\sc daophot} routines
\citep{setal90}. We generated a reference PSF, which was convolved
with the paw-print images to homogenize the resulting PSFs. We
repeated these steps for each epoch separately. Finally, all
homogenised paw-print images were combined using the {\sc swarp} tool
\citep{betal02}, as described in \cite{retal12}, thus generating deep
tile images with homogeneous PSFs.

We performed PSF photometry on the homogenized, deep tile images of
VMC tiles SMC 3$\_$5, 4$\_$5, 5$\_$6, 6$\_$5 and BRI 2$\_$3. 
We focussed on these tiles because they are among the first fully or mostly
completed tiles in the eastern region of the SMC (see column 4 of Table 1).
Figure 1 shows that they extend over a vast region on
the outermost, eastern side of the SMC and the associated onset of the
Bridge, where early galactic stellar populations and those recently
formed are supossed to co-exist as witnesses of the galaxy's formation
and its tidal interaction with the Large Magellanic Cloud (LMC). Note
that these tiles include, statistically speaking, an important
percentage of the star clusters catalogued in that region, thus
allowing us to draw robust conclusions about their star-formation
history.
The PSF model was created using $\sim$2500 stars which were
uniformly distributed and had magnitudes close to the saturation limit
+ 1.5 mag (for the VMC survey, the single paw-print saturation limits
are $Y=12.9$ mag, $J=12.7$ mag and $K_{\rm s} = 11.4$
mag). Subsequently, we used the {\sc allstar} routine to perform the
final PSF photometry on the deep tile images in all three filters and
correlated the resulting tables, adopting a tolerance of one
arcsec. We applied aperture corrections using catalogues retrieved
from the VSA\footnote{http://horus.roe.ac.uk/vsa/}
\citep{letal10b,cetal12} for the bulk of the observed stars.

We performed a large number of artificial-star tests to
estimate the incompleteness and error distribution of our data for
each deep tile image. For each VMC tile we generated $\sim 20 \times
10^6$ artificial stars following the steps outlined in \cite{retal12}. 
We used a
spatial grid of 25 pixels in width and with a spatial magnitude
distribution proportional to the square of the magnitude. This latter
choice allowed us to better map the completeness and error levels in
the less complete regions of the colour--magnitude diagram (CMD).
Table 1 lists the magnitudes of stars with photometric errors of less
than 0.10 mag (mag$_{\rm 0.1}$) and the magnitudes representing the
50\% completeness level. Our previous experience
\citep{cetal11,retal12,letal14} taught us that the widest colour
range, i.e., the $Y-K_{\rm s}$ colour, is optimal for cluster studies,
because it makes it easier to distinguish different cluster main
sequences (MSs), particularly their turn-off (TO) regions, and the
red-giant phases. This colour is also characterized by a higher
sensitivity to reddening and metallicity than either the $Y-J$ or
$J-K_{\rm s}$ colours. Consequently, the subsequent analysis is mainly
based on the $K_{\rm s}$ versus $Y-K_{\rm s}$ CMDs; the $J$ versus
$Y-J$ and $K_{\rm s}$ versus $J-K_{\rm s}$ CMDshave been constructed
and used to confirm the isochrones matched in the 
$K_{\rm s}$ versus $Y-K_{\rm s}$ CMDs.

\section{The cluster sample}

We delimited the outermost eastern region of the SMC and the
associated onset of the Bridge by an ellipse with a semimajor axis of
2.4$\degr$ and relative RA coordinates $\ge$ 1.7$\degr$ (see
Fig. 1). The star cluster population in that region comprises 79
objects according to the B08 catalogue; 27
(34\%) have been studied in previous photometric studies
\citep[see, e.g.,][]{getal10,pietal11}.  Fifty-one out of the 79
catalogued clusters are spread across the VMC tiles considered here,
which added to 12 other previously studied clusters located outside these
VMC tile areas, results in a more significant cluster sample (80\%)
than that previously studied. Note that the 51 clusters on the VMC tiles do
not only represent a larger percentage of the cluster population in
that region (65\%) with respect to the previously studied clusters,
but here we also analyse them in a homogeneous manner, thus enabling
us to trace their formation history spatially. To date, the catalogue 
of B08 is the most complete compilation of star clusters in the SMC
covering the extension of both Clouds and the Bridge. Particularly, because 
of the lower surface brightness of the background and the less-populous nature 
of the outer SMC regions and the Bridge, star clusters stand out more, 
and incompleteness effects are expected to be less important than
in the inner regions of the Clouds. 

Recognising catalogued star clusters in deep VMC tile images is
neither straighforward nor simple. On the one hand, the catalogued
objects were originally identified from optical images (e.g., from the
Digitized Sky Survey, DSS\footnote{The Digitized Sky Surveys were
  produced at the Space Telescope Science Institute under
  U.S. Government grant NAG W-2166. The images of these surveys are
  based on photographic data obtained using the Oschin Schmidt
  Telescope on Palomar Mountain and the UK Schmidt Telescope. The
  plates were processed into the present, compressed digital form with
  the permission of these institutions.}, images) which sometimes look
rather different compared with their appearance at near-infrared
wavelengths. In addition, the spatial resolution and depth of the
images on which the clusters were identified differ from the
equivalent parameters pertaining to the VMC images. Thus, for
instance, single relatively bright stars might look like an unresolved
compact cluster in images of lower spatial resolution, or unresolved
background galaxies could be mistaken for small star clusters in
shallower images. Offsets in the coordinates compiled by B08 with
respect to the objects' centres cannot be ruled out either.

To avoid mismatches between observed objects and the actual list of
catalogued clusters, we first overplotted the positions of the
catalogued clusters (B08) on the deepest $K_{\rm s}$ image. This way,
based on using the coordinates provided by B08, we visually recognized
the observed clusters one by one in the $K_{\rm s}$ image. Next, we
searched for these clusters in the DSS and downloaded
15$\arcmin$$\times$15$\arcmin$ $B$ images centred on the coordinates
resolved by the SIMBAD\footnote{http://simbad.u-strasbg.fr/simbad/}
astronomical data base and compared them with equivalent cut-outs
derived from the $K_{\rm s}$ VMC survey. Thus, we correctly recognized
the clusters in both the optical and near-infrared regimes. Note that
the main aim of this task is to confirm the cluster coordinates and
sizes, in order to extract from the tile-image PSF photometry the
magnitudes of the stars in the cluster region. We are not interested
in properties such as the clusters' structure, stellar density
profiles or radii, but in the stars which allow us to meaningfully
define the clusters' fiducial sequences in their CMDs. Table 2 lists
the complete cluster sample, as well as the coordinates and radii
adopted for extracting the stellar PSF photometry; the cluster's radius 
was taken either from a visual inspection of the deepest $K_s$ image
(where the profile disappears into the background noise), from B08
or from both sources combined. The observed
objects are of small angular size, typically $\sim$ 0.7$\arcmin$ in
diameter ($\sim$ 12.2 pc).

\section{Cleaning the cluster CMDs}

In general, the extracted cluster CMDs represent the composite stellar
populations distributed along the respective lines of sight. Since
they account for the luminosity function, colour distribution and
stellar density towards a given region on the sky, CMD analysis alone
might lead to an incorrect interpretation. On the other hand, the CMDs
of the stars located within a region around the catalogued cluster
centres are a helpful tool to assess the reality of the density
enhancements. They may imply that we are dealing with the presence of
a genuine star cluster, a chance grouping of stars along the line of
sight or a non-uniform distribution of interstellar material in the
region of interest. Note that catalogued cluster candidates appear on
the sky as small concentrations of stars, although this does not
necessarily imply that such concentrations constitute real, physically
bound systems.

For these reasons, we first statistically constructed CMDs representing
the field along the line of sight towards the
individual clusters, which we then used to clean the cluster CMDs. We
employed the cleaning procedure developed by \citet[see their
  fig. 12]{pb12}. The method compares the extracted cluster CMD to
four distinct CMDs composed of stars located reasonably far from the
object, but not too far so as to risk losing the local field-star
signature in terms of stellar density, luminosity function and/or
colour distribution. The four field regions were designed to cover an
equal area as that of the cluster (a circular area of three times the
cluster radius), and they were placed to the north, east, south and
west of the cluster area. Note that large areas were chosen to increase 
the statistics and hence the reliability of the cleaning process.

Comparisons of field and cluster CMDs have long been done by comparing
the numbers of stars counted in boxes distributed in a similar manner
throughout both CMDs. However, since some parts of the CMD are more
densely populated than others, counting the numbers of stars within
boxes of a fixed size is not universally efficient. For instance, to
deal with stochastic effects at relatively bright magnitudes (e.g.,
fluctuations in the numbers of bright stars), larger boxes are
required, while populous CMD regions can be characterized using
smaller boxes. Thus, use of boxes of different sizes distributed in
the same manner throughout both CMDs leads to a more meaningful
comparison of the numbers of stars in different CMD regions. By
starting with reasonably large boxes -- ($\Delta$$K_{\rm
  s},\Delta$($Y-K_{\rm s}$)) = (1.00, 0.25) mag -- centred on each
star in the four field CMDs and by subsequently reducing their sizes
until they reach the stars closest in magnitude and colour,
separately, we defined boxes which result in use of larger areas in
field CMD regions containing a small number of stars, and vice versa
(see Fig. 2, top right-hand panel, where we used an annulus 
-outer and inner radii equal to 3.163 and 3.0 times the cluster radius-
around the cluster instead of one of the four selected 
circular areas for illustrative purposes). 
Next, we plotted all these boxes
for each field CMD on the cluster CMD and subtracted the star located
closest to each box centre.

Since we repeated this task for each of the four field-CMD box
samples, we could assign a membership probability to each star in the
cluster CMD. This was done by counting the number of times a star
remained unsubtracted in the four cleaned cluster CMDs and by
subsequently dividing this number by four. Thus, we distinguished
field populations projected onto the cluster area, i.e., those stars
with a probability $P$ $\le$ 25\%, stars that could equally likely be
associated with either the field or the object of interest ($P$ =
50\%), and stars that are predominantly found in the cleaned cluster
CMDs ($P \ge$ 75\%) rather than in the field-star CMDs. Figure 2
illustrates the performance of the cleaning procedure for the HW77 field, where we
plotted three different CMDs: that for the stars located within the
cluster radius (top left-hand panel), a single-field CMD for an
annulus  -outer and inner radii equal to 3.163 and 3.0 times the cluster radius-
centred on the cluster, as well as the cleaned cluster
CMD (bottom left-hand panel). The schematic diagram with a superimposed 
circle of radius equal to the cluster {radius} is shown in the bottom 
right-hand panel. The pink,
light and dark blue solid circles in the bottom panels represent stars
with $P \le$ 25\%, $P = 50$\% and $P \ge$ 75$\%$, respectively. Note
that when comparing  observed cluster and field CMDs  for the HW\,77 region (top left- and right-hand 
panels), the cluster region
contains brighter red clump (RC) stars and a younger MSTO :
$K_{\rm s}$(RC$_{\rm HW77}$) $\sim$ 17.0 mag, $K_{\rm s}$(RC$_{\rm field}$) $\sim$ 17.5 mag,
$K_{\rm s}$(MSTO$_{\rm HW77}$) $\sim$ 19.5 mag and $K_{\rm s}$(MSTO$_{\rm field}$) $\sim$ 20.5 mag,
which suggests that HW77 is projected against a relatively older composite stellar population.
A full set of figures for the remaining objects studied here is provided in the
online version of the journal.

To assess whether a catalogued cluster is a  genuine aggregate, we inspected 
both the cluster's schematic diagram and the
cleaned CMD. Objects showing some apparent concentration  in the sky of stars with
 $P \ge$ 75\% that do not define clear
cluster sequences in the cleaned CMDs were  assumed to be possible
non-genuine aggregates. We found that nearly 30\% (15 catalogued
clusters) of our cluster sample fall into this latter category. This
relatively high percentage compares well with recent results of
different cluster sample analyses in both Magellanic Clouds, which
suggest that the B08 cluster catalogue might contain between 10\% and
30\% of  possible non-genuine aggregates \citep{pb12,p14,petal14b}. Likewise, we
found two clusters (GKH22 and GKH24) that could not be resolved by our
photometry and another (BS226) that could not be recognized in any
image using the coordinates given by B08. The relevant information
about the status of these objects is provided in the final column of
Table 2.

\section{Fundamental cluster parameters}

We used the cleaned cluster CMDs to estimate the fundamental cluster
parameters by matching the observations with the theoretical
isochrones of \citet{betal12}. In performing this task, one has to
deal with parameters such as the reddening, distance, age and
metallicity. Our strategy consisted of obtaining the reddening values
from an independent source, assuming the mean SMC distance modulus for
all clusters and adopting for the cluster ages the ages of the
isochrones which best reproduced their CMD features. We started with
isochrones for a metallicity of $Z = 0.003$ ([Fe/H] = $-0.7$ dex),
which corresponds to the mean SMC cluster metal content during the
last $\sim 2$--3 Gyr \citep{pg13}, and employed isochrones for other
metallicities as required. Note that the difference in $Y-K_{\rm s}$
colour between isochrones for $Z = 0.003$ and 0.001 ([Fe/H]= $-1.1$
dex) is $\sim$ 0.1 mag in the MSTO region, $\sim$ 0.05 mag at the
subgiant branch (SGB) and $\sim$ 0.15 mag at the tip of the red-giant
branch for ages between log($t$ yr$^{-1}$) = 9.3 (2 Gyr) and log($t$
yr$^{-1}$) = 9.6 (4.0 Gyr). This difference is $\sim$ 0.05 mag smaller
for log($t$ yr$^{-1}$) = 9.8 (6.3 Gyr).  This small difference suggests 
that the $Y-K_{\rm s}$ colour is not sensitive to metallicity
differences smaller than $\Delta$[Fe/H]=0.4 dex if we keep in mind the 
relatively sparse nature of the
majority of our clusters and the intrinsic spread of the stars in the
CMDs.  Indeed, we tried using isochrones with metallicities
[Fe/H] = -0.5 and -0.9 dex and found negligible differences with respect to
that of [Fe/H] = $-$0.7 dex. Nevertheless, only four clusters in our sample 
(L109, L110, B168
and HW85) appeared older than log($t$ yr$^{-1}$) = 9.3, and we
carefully paid attention to the effects of metallicity differences.

According to \citet{dutraetal01}, the SMC is optically thin,
characterized by average foreground and internal $E(B-V)$ colour
excesses of 0.05$\pm$0.05 mag and 0.04 mag (computed from a reddening
for the SMC surroundings of $E(B-V)$ = 0.02$\pm$0.02 mag), respectively. We took
advantage of the Magellanic Cloud extinction values based on RC
stellar photometry provided by the OGLE\,III collaboration
\citep{u03}, as described in \citet{hetal11}, to estimate
$E(B-V)_{OGLE\,III}$ colour excesses for 18 clusters in our sample
which are located in the area covered by OGLE\,III. Likewise, we made
use of the NASA/IPAC Extragalactic Data
base\footnote{http://ned.ipac.caltech.edu/. NED is operated by the Jet
  Propulsion Laboratory, California Institute of Technology, under
  contract with NASA.} (NED) to infer Galactic foreground reddening
values, $E(B-V)_{NED}$, for the same cluster list. Next, we computed
the difference, $\Delta E(B-V) = E(B-V)_{OGLE\,III}- E(B-V)_{NED} =
0.025 \pm 0.005$ mag, in good agreement with the mean internal
reddening given by \citet{dutraetal01}. Note that this mean SMC
internal reddening renders the isochrones 0.02 mag redder in $Y-K_{\rm
  s}$ and 0.01 mag fainter in $K_{\rm s}$, respectively, in the
cleaned cluster CMDs. In matching isochrones, we started by adopting
either $E(B-V)_{OGLE\,III}$ values or $E(B-V)_{NED}$ + 0.03 mag,
combined with the equations $E(Y-K_{\rm s}) = 0.84\times E(B-V)$ and
$K_{\rm s} = M_{K_{\rm s}} + (m-M)_0 + 0.372 \times E(B-V)$, assuming
$R_V = 3.1$ \citep{cetal89,getal13}. The final $E(B-V)$ values are
listed in Table 2. We found two clusters which may be affected by
differential reddening, i.e., HW74 \citep{betal04} and HW81
\citep[embedded in an H{\sc ii} region][]{tetal10c}. Two other
clusters (BS189, BS190) are characterized by noticeably large $E(B-V)$
 values, that may also be influenced by differential reddening. In 
particular, BS189 is by far the most reddened cluster known
in the Magellanic Clouds; it is even more heavily reddened than any
cluster in the highly reddened LMC/30 Doradus region \citep{tetal13}.

For all clusters, we adopted the mean SMC distance modulus $(m-M)_0$ =
18.90 $\pm$ 0.10 mag (60.0$^{+3.0}_{-2.5}$ kpc) \citep{getal10} and an
average line-of-sight SMC disc depth of 6 kpc \citep{cetal01}. Bearing
in mind that any cluster in the present sample could be placed in
front of or behind the SMC, we conclude that the difference in
apparent distance moduli could be as large as $\Delta({K_{\rm
    s}}-M_{K_{\rm s}})$ $\sim$ 0.2 mag. This difference is much
smaller than the difference between two closely spaced isochrones as
used here ($\Delta$ log($t$ yr$^{-1}$ = 0.1, $\Delta$ $M_{K_{\rm s}}$
$\sim$ 0.4 mag), so that adoption of a unique value for the distance
modulus does not dominate the final error budget incurred in matching
isochrones to the cluster CMDs.

In the matching procedure, we used subsets of isochrones ranging from
$\Delta$ log($t$ yr$^{-1}$) = $-0.3$ to +0.3, straddling the initial
rough age estimates. Since the \citet{betal12} isochrones are defined
in the Vegamag system (where Vega has a magnitude of zero in all
filters, by definition), we subtracted 0.074 mag in $Y$ and 0.003 mag
in $K_{\rm s}$ from the isochrones before matching them to the cluster
CMDs \citep{retal15}. Thus, we worked with both theoretical
models and observed CMDs which were defined in the VISTA system. We
found that isochrones bracketing the derived mean age by $\Delta$
log($t$ yr$^{-1}$) = $\pm$0.1 represent the overall age uncertainties
owing to the observed dispersion in the cluster CMDs. Although in some
cases the age dispersion is smaller than $\Delta$ log($t$ yr$^{-1}$) =
0.1, we prefer to keep the former value as an upper limit to our error
budget \citep[among others]{pietal11,p14b,petal14b}. Finally, we
adopted for the cluster age/metallicity those values corresponding to
the isochrone which best reproduced the cluster's main features in the
CMD. Table 2 lists the resulting age and metallicity values for 33 of
the confirmed clusters, while the bottom left-hand panel of Fig. 2
illustrates our isochrone-matching procedure in the context of the CMD
of HW77.

Fourteen clusters have previous age estimates based on matching of
theoretical isochrones to their CMDs. We previously studied seven of
these based on Washington $CT_1$ photometry and the remaining clusters
were analysed by \citet{getal10} as part of the Magellanic Cloud
Photometric Surveys \citep[MCPS][]{zetal02}. We took advantage of the
availability of the Washington data to perform a sound comparison of
the ages resulting from both the Washington photometry and the present
results. First, we cleaned the cluster CMDs following the precepts
described in Section 4. Then, we adopted the mean SMC distance
modulus, $E(B-V)$ colour excesses and metallicities discussed
above. Finally, we matched isochrones to the cluster CMDs, shifted by
$E(C-T_1) = 1.97 E(B-V)$ and $T_1$ = $M_{T_1} +2.62 E(B-V) + (m-M)_0$
\citep{gs99}. The resulting ages are listed in the final column of
Table 2. When comparing these values with the VMC cluster ages, we
found a difference of $\Delta$ log($t$ yr$^{-1}$) = $-$0.05$\pm$0.10,
with the VMC ages being slighlty older. On the other hand, a direct
comparison of the age estimates derived by \citet{getal10} (see the
final column of Table 2) and our values shows that our ages are
$\Delta$ log($t$ yr$^{-1}$) = $0.5 \pm 0.5$ older. We recall that
Glatt et al. did not perform any decontamination of field stars from
the cluster CMDs and they therefore had to decide whether to include
luminous supergiants in their fits, since the apparent MSTO of
sparse, young clusters is subject to pronounced statistical
fluctuations. It is possible that the lack of cleaned CMDs and their
shallower photometry did not allow them to achieve more reliable age
estimates.

\section{Analysis}

In this section we address the cluster-formation history on the
outermost, eastern side of the SMC and the connected onset of the
Bridge by analysing the spatial dependence of their age
distribution. We first enlarged the cluster sample by adding
previously studied objects located in this region but beyond the VMC
tile areas. We found a total of 12 clusters with available $CT_1$
Washington photometry and ages estimated from the matching of
isochrones to their CMDs. As shown above, these age estimates are in
very good agreement with those of the VMC clusters. We needed to
correct the $CT1_1$-based ages by only $\Delta$ log($t$ yr$^{-1}$) =
$-0.05$ to establish a homogeneous age scale, i.e., that of the
present VMC cluster ages. The properties of the additional cluster
sample are listed in Table 3.

The enlarged cluster sample is shown in Fig. 3 using solid circles. We
used a colour scale to distinguish among clusters with different
log($t$ yr$^{-1}$) values, so that darker colours correspond to older
clusters. The spatial distribution of the clusters seems to suggest
that while  the 8 oldest clusters -- log($t$ yr$^{-1}$) $\ge$ 9.6 
 (see Tables 2 and 3) -- are in
general located at greater distances to the galaxy's centre than the
younger clusters -- 9.0 $\le$ log($t$ yr$^{-1}$) $\le$ 9.4 -- there is
also a trail of young clusters, log($t$ yr$^{-1}$) $<$ 8.4, along the
onset of the Bridge. Since we are dealing with 80\% of the catalogued
clusters in this part of the SMC -- the remaining 20\% are mostly
located in the Bridge -- our results can be considered as based on an
unbiased cluster sample.

Viewing the SMC as a triaxial galaxy with the Dec, RA and line of
sight as the three axes, \citet{cetal01} found axial ratios of
approximately 1:2:4. Based on this result, and aiming at estimating
the clusters' projected distances to the SMC's centre, we adopted the
elliptical framework described in Section 2, where $a$ is the
semimajor axis of an ellipse centred on the SMC centre, for a $b/a$
axis ratio of 1/2 and one point of its trajectory coinciding with the
position of the cluster of interest. For our sample clusters with
log($t$ yr$^{-1}$) $\ge$ 9.6 we obtained $a$ = (6.0 $\pm$ 1.3)$\degr$,
in excellent agreement with the mean value for the 16 known SMC
clusters spanning the same age range \citep{detal01,p11c, dh98,
  metal98, petal07, p12b}. Note that both the present cluster sample
and the handful of the oldest known clusters span a semimajor axis
range between $\sim$ 2$\degr$ and 9$\degr$; the latter are found at
any position angle. For clusters with 9.0 $\le$ log($t$ yr$^{-1}$) $\le$
9.4, we obtained $a$ = (4.9 $\pm$ 1.8)$\degr$. Although the latter
result is statistically indistinguishable from that for the oldest
clusters in the sample, the smaller mean value suggests that the
intermediate-age clusters are spatially slightly more concentrated
than the older cluster population. This trend is in agreement with
\citet{cetal08}, \citet{p12a,p12b} and \citet{cetal13b}, who showed
that SMC clusters and field stars follow a similar trend in the sense
that the larger the semimajor axis is, the older their ages are, with
a non-negligible dispersion.

The other interesting feature of Fig. 3 is the trail of younger
clusters that seems to follow the Wing/Bridge components. Such a long
spatial sequence, which seems to originate directly in the inner SMC
disc, does not only harbour very young clusters, log($t$ yr$^{-1}$)
$\sim$ 7.3, but also some of intermediate ages, log($t$ yr$^{-1}$)
$\sim$ 9.1. \citet{metal05} imaged a star field located in the Wing
with the {\sl Hubble Space Telescope (HST)}'s WFPC2 camera and also
found a wide age range, i.e., log($t$ yr$^{-1}$) $\sim$ 8.0--9.2,
while \citet{setal09} used the {\sl HST}/ACS camera to observe two
fields in the Wing. They found evidence of enhancement in the
star-formation activity starting from log($t$ yr$^{-1}$) $\sim$
8.7. Continuous star-formation activity at intermediate ages, as well
as a very recent star-formation phase, have recently been found by
\citet{retal14} and \citet{setal14}. In addition, \citet{betal13} and
\citet{netal13} showed evidence of the presence of intermediate-age
stars that could have been stripped from the SMC, which has also been
predicted by models of gas replenishment from the SMC to the LMC
\citep{bch07}. \citet{h07} did not detect intermediate-age stellar
populations associated with the Bridge, which led him to conclude that
the material which was stripped from the SMC to form the Bridge may
have consisted of very nearly pure gas. However, the present cluster
ages agree well with results from field-star studies which showed that
the Bridge does not only contain gas from which new generations of
stars might be formed but also older stellar populations. Whether (i)
this star-formation process took place along the Wing/Bridge
components, (ii) clusters/field stars were stripped from the SMC in
the direction of the LMC and (iii) their formation history is a
continuous or a burst-like chain of events are questions that are
still being debated and which deserve further investigation. From the
perspective of the clusters' ages, the Wing/Bridge structures seem to
have existed at least during the past 1--2 Gyr, i.e., log($t$
yr$^{-1}$) $\sim$ 9.0--9.3.

Figure 4 shows a time series pertaining to the clusters' spatial
distribution in intervals of $\Delta$log($t$ yr$^{-1}$) = 0.2; the lower limit
included in the interval. We have
drawn, in every panel, all catalogued clusters (dots) and the same two
ellipses as in Fig. 3. The present cluster sample (Tables 2 and 3) is
shown as solid circles. Figure 4 reveals that the ages of the clusters
in the outermost, eastern part of the SMC cover the entire age range,
with the exception of the oldest ages and the interval 7.4 $\le$ log($t$
yr$^{-1}$) $<$ 7.8. The lack of SMC clusters that are as old as the
Galactic globular clusters has long been known \citep{getal08b}. 
On the other hand, the absence of clusters in the
younger age range indicated can be linked to the minimum age found by
\citet{getal10}, log($t$ yr$^{-1}$) $\sim$ 7.5, based on an age
distribution composed of 821 clusters and associations distributed
across the main body of the galaxy; only ages with $\Delta$log($t$
yr$^{-1}$) $<$ 0.5 were used.

Clusters older than log($t$ yr$^{-1}$) $\sim$ 9.2 appear to be found
predominantly in the outer disc, whereas younger clusters tend to be
mostly located along the Wing/Bridge interface. Two peaks populated by
older clusters are seen at log($t$ yr$^{-1}$) $\sim$ 9.0--9.4 and
log($t$ yr$^{-1}$) $\sim$ 9.6--9.8, in very good agreement with
enhanced episodes of cluster formation that occurred throughout the
galaxy \citet[and reference therein]{p11b}. Both peaks were found by
\citet{netal09} from the recovery of the formation history of stars
observed in 12 fields in the $B,R$ bands with the 100 inch telescope
at Las Campanas Observatory. \citet{hz04} also detected the younger
peak based on their MCPS data, whereas \citet{retal14b} showed the
presence of the older peak based on observations of long-period
variable stars.

\subsection{Star cluster frequency}

We constructed the star cluster  formation frequency as a function of 
age (CF) -- the number of
clusters per unit time interval as a function of age -- to compare it
with that derived under the assumption that they are characterized by
a similar formation history as that of the field stars projected along
their respective lines of sight. We were interested in obtaining the
CFs for clusters located in tiles SMC 3$\_$5, 4$\_$5,
5$\_$6 and 6$\_$5 for which \citet{retal15} obtained star-formation rates using
the same VMC photometric data set.

The first step in generating the CF consisted of considering the age
errors. Indeed, by taking into account such errors, the interpretation
of the resulting CF can differ appreciably from that obtained using
only the measured ages without accounting for their
uncertainties. However, treatment of the age errors in the CF is not
straightforward. Moreover, even if errors did not play an important
role, binning of the age ranges could also bias the results. At first
glance, a fixed age-bin size may not be appropriate, since the
resulting age distribution depends on the adopted age interval, and
the age errors are typically a strong function of age. On the
contrary, adoption of age-bin widths on the order of the clusters' age
errors in the interval of interest appears more meaningful. This would
lead to the selection of narrow bins (in linear age) for young
clusters and relatively broader age bins at older ages
\citep{pg13,p14}. To account for the effects of the age uncertainties
in the CF while taking into account that any individual point in the
CF may fall either in the respective age bin or in any of the adjacent
bins, we followed \citet{p14b}.

For our purposes, we first considered the cluster age range, split
into bins of $\Delta$ log($t$ yr$^{-1}$) = 0.10. On the other hand,
each age data point and its associated uncertainty ($\sigma(t)$)
covers a segment with a size of 2$\times$$\sigma(t)$ (which may or may
not fall fully in one of the age bins) and has dimensions smaller,
similar to or larger than the age bin in which it is found. For this
reason, we weighted the contribution of each age data point to each of
the associated age bins, with the sum total of all weights being equal
to unity. The assigned weight was computed as the fraction of the age
segment that falls within the relevant age bin. In practice, for each
age interval we looked for clusters with ages that fall inside the age
bin considered, as well as clusters where $t$ $\pm$ $\sigma(t)$ could
cause them to fall in the same age bin. Note that if $t$ $\pm$
$\sigma(t)$ causes a cluster's age range to extend beyond the main age
bin considered (e.g., from one bin to the next), then we consider that
that cluster may have an age that also places it inside the extended
age interval considered. Figure 5 shows the resulting CF (thick solid
line). This CF has been normalized to the total number of
clusters. Note the discontinuity at log($t$ yr$^{-1}$) = 7.4--7.8,
which may be owing to the small number of clusters in our sample. The
younger part of the CF is mainly based on clusters located in the
inner tiles, SMC 3$\_$5 and 4$\_$5, while the remaining outer tiles
(5$\_$6 and SMC 6$\_$5) mostly contribute to the older regime.

To place the CF resulting from our small cluster sample in the overall
context of the SMC's cluster population, we also included the CF
constructed on the basis of the cluster age determinations of de Grijs
\& Goodwin (2008; thick dashed line). The latter authors obtained
best-fitting ages for more than 300 SMC clusters based on broad-band
photometric measurements at optical wavelengths. They validated their
resulting age estimates with respect to both those obtained by other
teams using similar approaches and by reference to their extensive
validation of the modelling approach used. Despite the differences in
both cluster numbers and photometric completeness limits between both
studies, the overall trends seen in Fig. 5 are similar. 

The effects of small-number statistics are apparent in the stochastic
fluctuations in the CF based on the VMC data set. The main difference
between the CF constructed based on the VMC data set with respect to
that resulting from the de Grijs \& Goodwin (2008) data is the
relatively larger number of older clusters in the VMC data. A careful
comparison between both data sets reveals that the completeness limit
in terms of cluster masses of the VMC data sets in at much lower
masses than that pertaining to the broad-band data set. The older,
lower-mass clusters seen in the near-infrared VMC data had already
faded to below the detection limit in the bluest filter used by de
Grijs \& Anders (2008). All other apparent differences between both
observational CFs can be traced back to small-number statistics. Note
that the similarity between both CFs, which are nevertheless based on
very different lower-mass limits, implies indirectly that the cluster
mass function in the SMC is well represented by a power-law function
down to the lowest accessible masses (cf. de Grijs \& Goodwin 2008).

Finally, we compare the present, VMC-based CF to that obtained from
the star-formation rates derived by \citet{retal15} from the same
VMC tiles and PSF photometry data set used here. They subdivided
each tile into 12 subregions of 21.0$\arcmin$$\times$21.5$\arcmin$
($\sim$ 0.12 deg$^2$) and derived their star-formation histories as
described in detail in \citet{retal12}. Briefly, for a range of
distance moduli and visual extinction estimates, `partial models' are
derived for the entire range of metallicities and ages of
relevance. Partial models are synthetic stellar populations, each
covering small bins in age and metallicity, shifted to the desired
distance and extinction. Each partial model is `degraded' to the
conditions of the actual observations by convolving them with the
distributions of photometric errors and completeness. The linear
combination of partial models that optimally matches the observed Hess
diagrams is found by means of the StarFISH optimisation code
\citep{hz01}. The coefficients of this linear combination of partial
models (including the best-fitting distance modulus and extinction)
are directly converted into the star-formation rates.

We assumed that clusters are formed with a power-law mass distribution
characterized by a slope of $\alpha = -2$ \citep{baetal13,p14b} and
with a rate that is proportional to the field-star formation rate
determined by \citet{retal15} for the individual subfields pertaining
to our clusters, taking into account the corresponding uncertainties.
We used cluster masses from log($M_{\rm cl}$ [M$_\odot$]) = 2.2 to
log($M_{\rm cl}$ [M$_\odot$]) = 5.0 \citep{dgetal08,getal11} and
normalized the resulting CF by the total number of clusters used, so
that it can be compared directly to the observed distribution. Figure
5 shows the resulting model CF, where the uncertainties are indicated
using thin solid lines. As can be seen, the observed and modeled CFs 
are different. For ages in excess of log($t$ yr$^{-1}$) $\sim$ 9.4
the model CF is higher than the observed frequency, possibly because
we did not take into account any cluster dissolution. According to
\citet{ll03}, most -- if not all -- stars form in some sort of
cluster. This implies that field stars are the result of cluster
dissolution and do not originate from an independent formation
mechanism, so that the difference between the observed and modeled CFs
would then be the result of the prevailing cluster dissolution rate.

The decrease of the star-formation activity that started at log($t$
yr$^{-1}$) $\sim$ 9.0 and was interrupted by a burst-like formation
event which occurred at log($t$ yr$^{-1}$) $\sim$ 8.5 was also
documented by \citet{hz04}, \citet{netal09}, \citet{setal09} and
\citet{retal15}. However, the observed CF is significantly higher for
ages younger than log($t$ yr$^{-1}$) $\sim$ 9.0. Even though the model
CF requires additional refinements, the clear cluster excess deserves
further analysis. If we now keep in mind \citet{dgetal13}'s results,
who showed that there is no evidence of significant destruction other
than that expected from stellar dynamics and evolution for simple
stellar population models for ages up to 1 Gyr -- log($t$ yr$^{-1}$) =
9 -- we speculate that such a cluster excess could be evidence of
enhanced cluster formation in this part of the SMC with respect to
that on the western side. Indeed, the number of clusters on that
latter side is significantly smaller (see Fig. 3). Assuming that there
probably has been no significant mixing of clusters among different
regions, the observed excess of clusters reveals that the cluster
formation history in the outermost, eastern region of the SMC could
have been influenced at some level by the galaxy's interaction with
the LMC.  Such an interaction \citep{netal13b,retal14b} could have resulted 
in either by stripping mechanics \citep{betal13,netal13}, or in situ star formation 
processes \citep{metal05,retal14}, or a combination of both that caused
the observed and modeled CFs to look different for ages younger than
 1 Gyr.

\section{Conclusions}

In this paper, we have analysed CMDs of catalogued star clusters
located in the outermost, eastern region of the SMC based on a
$YJK_{\rm s}$ photometric data set obtained by the VISTA VMC
collaboration. We focussed on tiles SMC 3$\_$5, 4$\_$5, 5$\_$6, 6$\_$5,
and BRI 2$\_$3, because they are among the first fully or mostly
completed tiles in the eastern region of the SMC, which may also show
evidence of the SMC's interaction history with the LMC. We obtained
PSF photometry for stars in and projected towards 51 catalogued
clusters of small angular size, typically $\sim$ 12.2 pc in diameter,
which represents a meaningful sample size for scientific studies.

We applied a field-star subtraction procedure to statistically clean
the cluster CMDs from field-star contamination to disentangle cluster
features from those associated with their surrounding fields. The
technique we employed makes use of variable cells to reproduce the
field CMDs as closely as possible. We found that nearly 30\% (15
catalogued clusters) of the cluster sample investigated
do not have cluster sequences in the cleaned CMDs an were therefore
assumed to be possible non-genuine aggregates.

Based on matching theoretical isochrones in the VISTA system to the
cleaned cluster CMDs, we obtained estimates of the cluster ages,
assuming as initial guess for the cluster metalicity the value $Z =
0.003$. We took into account the SMC's distance modulus as well as the
individual cluster colour excesses. The resulting cluster ages span
the age range 7.0 $\le$ log($t$ yr$^{-1}$) $\le$ 9.8. This cluster sample
forms part of the cluster data base which will result from the VMC
survey and which will be used to self-consistently study the overall
cluster formation history of the Magellanic system. Fourteen clusters
have previous age estimates based on the matching of isochrones to
their CMDs. We studied seven of these based on Washington $CT_1$
photometry, whereas the remaining clusters were studied by
\citet{getal10}. Although the present cluster ages are in very good
agreement with those resulting from analysis of Washington photometry,
\citet{getal10}'s ages are much younger, possibly because of the lack
of cleaned CMDs and the shallower photometry used by these authors.

We complemented the properties of 80\% of the catalogued clusters in
this part of the SMC by increasing our sample by addition of
previously determined cluster ages, taking care to adopt a homogeneous
scale. Their spatial distribution seems to suggest that the oldest
clusters -- log($t$ yr$^{-1}$) $\le$ 9.6 -- are in general located at
greater distances from the galaxy's centre than their younger
counterparts, 9.0 $\le$ log($t$ yr$^{-1}$) $\le$ 9.4, which is in
agreement with previous studies of star clusters and field stars. The
latter showed that the greater their distances to the SMC centre, the
older the clusters' ages are, with a non-negligible dispersion. Within
the older cluster population, two excesses of clusters are seen, at
log($t$ yr$^{-1}$) $\sim$ 9.0--9.4 and log($t$ yr$^{-1}$) $\sim$
9.6--9.8, in very good agreement with the timing of enhanced episodes
of cluster formation previously found throughout the galaxy.

We also found a trail of younger clusters that seems to follow the
Wing/Bridge components. Such a long spatial sequence, which extends
from the inner SMC disc, does not only harbour very young clusters --
log($t$ yr$^{-1}$) $\sim$ 7.3 -- but it also includes some of
intermediate ages, log($t$ yr$^{-1}$) $\sim$ 9.1. Therefore, the
Wing/Bridge structures seem to have existed at least during the last
1--2 Gyr, log($t$ yr$^{-1}$).

The observed  and modeled CF, excluding tile BRI 2$\_$3, 
 are different. For ages in excess of log($t$ yr$^{-1}$)
$\sim$ 9.4, the model CF, obtained by assuming a cluster formation
rate proportional to that of the field stars and a power-law mass
distribution with a slope of $\alpha$ = $-$2, is higher than the
observed frequency, possibly because we did not take into account any
cluster dissolution. For younger ages, the observed CF is --
surprisingly -- significantly higher than the model distribution,
which may be partially owing to the fact that field-star formation
activity is known to decrease from log($t$ yr$^{-1}$) $\sim$ 9.0 until
a burst-like formation event occurred at log($t$ yr$^{-1}$) $\sim$
8.5. Nevertheless, even though the model CF requires additional
refinements, if we assume that there probably has been no significant
mixing of clusters among different regions, the observed excess of
clusters reveals that the cluster formation history in the outermost,
eastern region of the SMC could have been influenced at some level by
the galaxy's interaction history with the LMC.

\section*{Acknowledgements}
We thank the Cambridge Astronomy Survey Unit (CASU) and the Wide-Field
Astronomy Unit (WFAU) in Edinburgh for providing calibrated data
products under the support of the Science and Technology Facilities
Council (STFC) in the UK. This research has made use of the SIMBAD
data base, operated at CDS, Strasbourg, France, and draws upon data
distributed through the NOAO Science Archive. We thank the anonymous 
referee whose 
comments and suggestions allowed us to improve the manuscript.
This work was partially supported by the Argentinian institutions
CONICET and Agencia Nacional de Promoci\'on Cient\'{\i}fica y
Tecnol\'ogica (ANPCyT). RdG acknowledges research support from the
National Natural Science Foundation of China (NSFC; grant 11373010).

\bibliographystyle{mn2e_new} 
\bibliography{paper} 

\begin{thebibliography}{67}
\expandafter\ifx\csname natexlab\endcsname\relax\def\natexlab#1{#1}\fi

\bibitem[{{Arnaboldi} {et~al}\mbox{.}(2013){Arnaboldi}, {Rejkuba}, {Retzlaff},
  {Delmotte}, {Geier}, {Hanuschik}, {Hilker}, {Hummel}, {Hussain}, {Ivanov},
  {Micol}, {Mieske}, {Neeser}, {Petr-Gotzens}, \& {Szeifert}}]{aetal13}
{Arnaboldi} M. {et~al.}, 2013, The Messenger, 154, 18

\bibitem[{{Bagheri}, {Cioni} \& {Napiwotzki}(2013){Bagheri}, {Cioni}, \&
  {Napiwotzki}}]{betal13}
{Bagheri} G., {Cioni} M.-R.~L., {Napiwotzki} R., 2013, \aap, 551, A78

\bibitem[{{Baumgardt} {et~al}\mbox{.}(2013){Baumgardt}, {Parmentier}, {Anders},
  \& {Grebel}}]{baetal13}
{Baumgardt} H., {Parmentier} G., {Anders} P., {Grebel} E.~K., 2013, \mnras,
  430, 676

\bibitem[{{Bekki} \& {Chiba}(2007)}]{bch07}
{Bekki} K., {Chiba} M., 2007, \mnras, 381, L16

\bibitem[{{Bertin} {et~al}\mbox{.}(2002){Bertin}, {Mellier}, {Radovich},
  {Missonnier}, {Didelon}, \& {Morin}}]{betal02}
{Bertin} E., {Mellier} Y., {Radovich} M., {Missonnier} G., {Didelon} P.,
  {Morin} B., 2002, in Astronomical Society of the Pacific Conference Series,
  Vol. 281, Astronomical Data Analysis Software and Systems XI,
  {D.~A.~Bohlender, D.~Durand, \& T.~H.~Handley}, ed., pp. 228--+

\bibitem[{{Bica} {et~al}\mbox{.}(2008){Bica}, {Bonatto}, {Dutra}, \&
  {Santos}}]{betal08}
{Bica} E., {Bonatto} C., {Dutra} C.~M., {Santos} J.~F.~C., 2008, \mnras, 389,
  678

\bibitem[{{Bica}, {Santos} \& {Schmidt}(2008){Bica}, {Santos}, \&
  {Schmidt}}]{betal08b}
{Bica} E., {Santos}, Jr. J.~F.~C., {Schmidt} A.~A., 2008, \mnras, 391, 915

\bibitem[{{Bratsolis}, {Kontizas} \& {Bellas-Velidis}(2004){Bratsolis},
  {Kontizas}, \& {Bellas-Velidis}}]{betal04}
{Bratsolis} E., {Kontizas} M., {Bellas-Velidis} I., 2004, \aap, 423, 919

\bibitem[{{Bressan} {et~al}\mbox{.}(2012){Bressan}, {Marigo}, {Girardi},
  {Salasnich}, {Dal Cero}, {Rubele}, \& {Nanni}}]{betal12}
{Bressan} A., {Marigo} P., {Girardi} L., {Salasnich} B., {Dal Cero} C.,
  {Rubele} S., {Nanni} A., 2012, \mnras, 427, 127

\bibitem[{{Cardelli}, {Clayton} \& {Mathis}(1989){Cardelli}, {Clayton}, \&
  {Mathis}}]{cetal89}
{Cardelli} J.~A., {Clayton} G.~C., {Mathis} J.~S., 1989, \apj, 345, 245

\bibitem[{{Carrera} {et~al}\mbox{.}(2008){Carrera}, {Gallart}, {Aparicio},
  {Costa}, {M{\'e}ndez}, \& {No{\"e}l}}]{cetal08}
{Carrera} R., {Gallart} C., {Aparicio} A., {Costa} E., {M{\'e}ndez} R.~A.,
  {No{\"e}l} N.~E.~D., 2008, \aj, 136, 1039

\bibitem[{{Cignoni} {et~al}\mbox{.}(2013){Cignoni}, {Cole}, {Tosi},
  {Gallagher}, {Sabbi}, {Anderson}, {Grebel}, \& {Nota}}]{cetal13b}
{Cignoni} M., {Cole} A.~A., {Tosi} M., {Gallagher} J.~S., {Sabbi} E.,
  {Anderson} J., {Grebel} E.~K., {Nota} A., 2013, \apj, 775, 83

\bibitem[{{Cioni} {et~al}\mbox{.}(2011){Cioni}, {Clementini}, {Girardi},
  {Guandalini}, {Gullieuszik}, {Miszalski}, {Moretti}, {Ripepi}, {Rubele},
  {Bagheri}, {Bekki}, {Cross}, {de Blok}, {de Grijs}, {Emerson}, {Evans},
  {Gibson}, {Gonzales-Solares}, {Groenewegen}, {Irwin}, {Ivanov}, {Lewis},
  {Marconi}, {Marquette}, {Mastropietro}, {Moore}, {Napiwotzki}, {Naylor},
  {Oliveira}, {Read}, {Sutorius}, {van Loon}, {Wilkinson}, \& {Wood}}]{cetal11}
{Cioni} M.-R.~L. {et~al.}, 2011, \aap, 527, A116

\bibitem[{{Cross} {et~al}\mbox{.}(2012){Cross}, {Collins}, {Mann}, {Read},
  {Sutorius}, {Blake}, {Holliman}, {Hambly}, {Emerson}, {Lawrence}, \&
  {Noddle}}]{cetal12}
{Cross} N.~J.~G. {et~al.}, 2012, \aap, 548, A119

\bibitem[{{Crowl} {et~al}\mbox{.}(2001){Crowl}, {Sarajedini}, {Piatti},
  {Geisler}, {Bica}, {Clari{\'a}}, \& {Santos}}]{cetal01}
{Crowl} H.~H., {Sarajedini} A., {Piatti} A.~E., {Geisler} D., {Bica} E.,
  {Clari{\'a}} J.~J., {Santos}, Jr. J.~F.~C., 2001, \aj, 122, 220

\bibitem[{{Da Costa} \& {Hatzidimitriou}(1998)}]{dh98}
{Da Costa} G.~S., {Hatzidimitriou} D., 1998, \aj, 115, 1934

\bibitem[{{de Grijs} \& {Goodwin}(2008)}]{dgetal08}
{de Grijs} R., {Goodwin} S.~P., 2008, \mnras, 383, 1000

\bibitem[{{de Grijs}, {Goodwin} \& {Anders}(2013){de Grijs}, {Goodwin}, \&
  {Anders}}]{dgetal13}
{de Grijs} R., {Goodwin} S.~P., {Anders} P., 2013, \mnras, 436, 136

\bibitem[{{Dolphin} {et~al}\mbox{.}(2001){Dolphin}, {Walker}, {Hodge}, {Mateo},
  {Olszewski}, {Schommer}, \& {Suntzeff}}]{detal01}
{Dolphin} A.~E., {Walker} A.~R., {Hodge} P.~W., {Mateo} M., {Olszewski} E.~W.,
  {Schommer} R.~A., {Suntzeff} N.~B., 2001, \apj, 562, 303

\bibitem[{{Dutra} {et~al}\mbox{.}(2001){Dutra}, {Bica}, {Clari{\'a}}, {Piatti},
  \& {Ahumada}}]{dutraetal01}
{Dutra} C.~M., {Bica} E., {Clari{\'a}} J.~J., {Piatti} A.~E., {Ahumada} A.~V.,
  2001, \aap, 371, 895

\bibitem[{{Gao} {et~al}\mbox{.}(2013){Gao}, {Jiang}, {Li}, \& {Xue}}]{getal13}
{Gao} J., {Jiang} B.~W., {Li} A., {Xue} M.~Y., 2013, \apj, 776, 7

\bibitem[{{Geisler} \& {Sarajedini}(1999)}]{gs99}
{Geisler} D., {Sarajedini} A., 1999, \aj, 117, 308

\bibitem[{{Glatt} {et~al}\mbox{.}(2008){Glatt}, {Gallagher}, {Grebel}, {Nota},
  {Sabbi}, {Sirianni}, {Clementini}, {Tosi}, {Harbeck}, {Koch}, \&
  {Cracraft}}]{getal08b}
{Glatt} K. {et~al.}, 2008, \aj, 135, 1106

\bibitem[{{Glatt} {et~al}\mbox{.}(2011){Glatt}, {Grebel}, {Jordi}, {Gallagher},
  {Da Costa}, {Clementini}, {Tosi}, {Harbeck}, {Nota}, {Sabbi}, \&
  {Sirianni}}]{getal11}
---, 2011, \aj, 142, 36

\bibitem[{{Glatt}, {Grebel} \& {Koch}(2010){Glatt}, {Grebel}, \&
  {Koch}}]{getal10}
{Glatt} K., {Grebel} E.~K., {Koch} A., 2010, \aap, 517, A50

\bibitem[{{Gouliermis}, {Quanz} \& {Henning}(2007){Gouliermis}, {Quanz}, \&
  {Henning}}]{getal07}
{Gouliermis} D.~A., {Quanz} S.~P., {Henning} T., 2007, \apj, 665, 306

\bibitem[{{Harris}(2007)}]{h07}
{Harris} J., 2007, \apj, 658, 345

\bibitem[{{Harris} \& {Zaritsky}(2001)}]{hz01}
{Harris} J., {Zaritsky} D., 2001, \apjs, 136, 25

\bibitem[{{Harris} \& {Zaritsky}(2004)}]{hz04}
---, 2004, \aj, 127, 1531

\bibitem[{{Haschke}, {Grebel} \& {Duffau}(2011){Haschke}, {Grebel}, \&
  {Duffau}}]{hetal11}
{Haschke} R., {Grebel} E.~K., {Duffau} S., 2011, \aj, 141, 158

\bibitem[{{Irwin} {et~al}\mbox{.}(2004){Irwin}, {Lewis}, {Hodgkin}, {Bunclark},
  {Evans}, {McMahon}, {Emerson}, {Stewart}, \& {Beard}}]{ietal04}
{Irwin} M.~J. {et~al.}, 2004, in Society of Photo-Optical Instrumentation
  Engineers (SPIE) Conference Series, Vol. 5493, Optimizing Scientific Return
  for Astronomy through Information Technologies, {Quinn} P.~J., {Bridger} A.,
  eds., pp. 411--422

\bibitem[{{Lada} \& {Lada}(2003)}]{ll03}
{Lada} C.~J., {Lada} E.~A., 2003, \araa, 41, 57

\bibitem[{{Lewis}, {Irwin} \& {Bunclark}(2010){Lewis}, {Irwin}, \&
  {Bunclark}}]{letal10b}
{Lewis} J.~R., {Irwin} M., {Bunclark} P., 2010, in Astronomical Society of the
  Pacific Conference Series, Vol. 434, Astronomical Data Analysis Software and
  Systems XIX, {Mizumoto} Y., {Morita} K.-I., {Ohishi} M., eds., p.~91

\bibitem[{{Li} {et~al}\mbox{.}(2014){Li}, {de Grijs}, {Deng}, {Rubele}, {Wang},
  {Bekki}, {Cioni}, {Clementini}, {Emerson}, {For}, {Girardi}, {Groenewegen},
  {Guandalini}, {Gullieuszik}, {Marconi}, {Piatti}, {Ripepi}, \& {van
  Loon}}]{letal14}
{Li} C. {et~al.}, 2014, \apj, 790, 35

\bibitem[{{McCumber}, {Garnett} \& {Dufour}(2005){McCumber}, {Garnett}, \&
  {Dufour}}]{metal05}
{McCumber} M.~P., {Garnett} D.~R., {Dufour} R.~J., 2005, \aj, 130, 1083

\bibitem[{{Mighell}, {Sarajedini} \& {French}(1998){Mighell}, {Sarajedini}, \&
  {French}}]{metal98}
{Mighell} K.~J., {Sarajedini} A., {French} R.~S., 1998, \aj, 116, 2395

\bibitem[{{Moretti} {et~al}\mbox{.}(2014){Moretti}, {Clementini}, {Muraveva},
  {Ripepi}, {Marquette}, {Cioni}, {Marconi}, {Girardi}, {Rubele}, {Tisserand},
  {de Grijs}, {Groenewegen}, {Guandalini}, {Ivanov}, \& {van Loon}}]{metal14c}
{Moretti} M.~I. {et~al.}, 2014, \mnras, 437, 2702

\bibitem[{{Nidever} {et~al}\mbox{.}(2013){Nidever}, {Monachesi}, {Bell},
  {Majewski}, {Mu{\~n}oz}, \& {Beaton}}]{netal13b}
{Nidever} D.~L., {Monachesi} A., {Bell} E.~F., {Majewski} S.~R., {Mu{\~n}oz}
  R.~R., {Beaton} R.~L., 2013, \apj, 779, 145

\bibitem[{{No{\"e}l} {et~al}\mbox{.}(2009){No{\"e}l}, {Aparicio}, {Gallart},
  {Hidalgo}, {Costa}, \& {M{\'e}ndez}}]{netal09}
{No{\"e}l} N.~E.~D., {Aparicio} A., {Gallart} C., {Hidalgo} S.~L., {Costa} E.,
  {M{\'e}ndez} R.~A., 2009, \apj, 705, 1260

\bibitem[{{No{\"e}l} {et~al}\mbox{.}(2013){No{\"e}l}, {Conn}, {Carrera},
  {Read}, {Rix}, \& {Dolphin}}]{netal13}
{No{\"e}l} N.~E.~D., {Conn} B.~C., {Carrera} R., {Read} J.~I., {Rix} H.-W.,
  {Dolphin} A., 2013, \apj, 768, 109

\bibitem[{{Parisi} {et~al}\mbox{.}(2014){Parisi}, {Geisler}, {Carraro},
  {Clari{\'a}}, {Costa}, {Grocholski}, {Sarajedini}, {Leiton}, \&
  {Piatti}}]{paetal14}
{Parisi} M.~C. {et~al.}, 2014, \aj, 147, 71

\bibitem[{{Piatti}(2011{\natexlab{a}})}]{p11c}
{Piatti} A.~E., 2011{\natexlab{a}}, \mnras, 416, L89

\bibitem[{{Piatti}(2011{\natexlab{b}})}]{p11b}
---, 2011{\natexlab{b}}, \mnras, 418, L69

\bibitem[{{Piatti}(2012{\natexlab{a}})}]{p12b}
---, 2012{\natexlab{a}}, \apjl, 756, L32

\bibitem[{{Piatti}(2012{\natexlab{b}})}]{p12a}
---, 2012{\natexlab{b}}, \mnras, 422, 1109

\bibitem[{{Piatti}(2014{\natexlab{a}})}]{p14}
---, 2014{\natexlab{a}}, \mnras, 440, 3091

\bibitem[{{Piatti}(2014{\natexlab{b}})}]{p14b}
---, 2014{\natexlab{b}}, \mnras, 437, 1646

\bibitem[{{Piatti} \& {Bica}(2012)}]{pb12}
{Piatti} A.~E., {Bica} E., 2012, \mnras, 425, 3085

\bibitem[{{Piatti} {et~al}\mbox{.}(2011){Piatti}, {Clari{\'a}}, {Bica},
  {Geisler}, {Ahumada}, \& {Girardi}}]{pietal11}
{Piatti} A.~E., {Clari{\'a}} J.~J., {Bica} E., {Geisler} D., {Ahumada} A.~V.,
  {Girardi} L., 2011, \mnras, 417, 1559

\bibitem[{{Piatti} \& {Geisler}(2013)}]{pg13}
{Piatti} A.~E., {Geisler} D., 2013, \aj, 145, 17

\bibitem[{{Piatti} {et~al}\mbox{.}(2008){Piatti}, {Geisler}, {Sarajedini},
  {Gallart}, \& {Wischnjewsky}}]{pietal08}
{Piatti} A.~E., {Geisler} D., {Sarajedini} A., {Gallart} C., {Wischnjewsky} M.,
  2008, \mnras, 389, 429

\bibitem[{{Piatti} {et~al}\mbox{.}(2014){Piatti}, {Guandalini}, {Ivanov},
  {Rubele}, {Cioni}, {de Grijs}, {For}, {Clementini}, {Ripepi}, {Anders}, \&
  {Oliveira}}]{petal14b}
{Piatti} A.~E. {et~al.}, 2014, ArXiv e-prints

\bibitem[{{Piatti} {et~al}\mbox{.}(2007{\natexlab{a}}){Piatti}, {Sarajedini},
  {Geisler}, {Clark}, \& {Seguel}}]{petal07d}
{Piatti} A.~E., {Sarajedini} A., {Geisler} D., {Clark} D., {Seguel} J.,
  2007{\natexlab{a}}, \mnras, 377, 300

\bibitem[{{Piatti} {et~al}\mbox{.}(2007{\natexlab{b}}){Piatti}, {Sarajedini},
  {Geisler}, {Gallart}, \& {Wischnjewsky}}]{petal07b}
{Piatti} A.~E., {Sarajedini} A., {Geisler} D., {Gallart} C., {Wischnjewsky} M.,
  2007{\natexlab{b}}, \mnras, 382, 1203

\bibitem[{{Piatti} {et~al}\mbox{.}(2007{\natexlab{c}}){Piatti}, {Sarajedini},
  {Geisler}, {Gallart}, \& {Wischnjewsky}}]{petal07}
---, 2007{\natexlab{c}}, \mnras, 381, L84

\bibitem[{{Rezaei kh.} {et~al}\mbox{.}(2014){Rezaei kh.}, {Javadi},
  {Khosroshahi}, \& {van Loon}}]{retal14b}
{Rezaei kh.} S., {Javadi} A., {Khosroshahi} H., {van Loon} J.~T., 2014, ArXiv
  e-prints

\bibitem[{{Ripepi} {et~al}\mbox{.}(2014){Ripepi}, {Cignoni}, {Tosi}, {Marconi},
  {Musella}, {Grado}, {Limatola}, {Clementini}, {Brocato}, {Cantiello},
  {Capaccioli}, {Cappellaro}, {Cioni}, {Cusano}, {Dall'Ora}, {Gallagher},
  {Grebel}, {Nota}, {Palla}, {Romano}, {Raimondo}, {Sabbi}, {Getman},
  {Napolitano}, {Schipani}, \& {Zaggia}}]{retal14}
{Ripepi} V. {et~al.}, 2014, \mnras, 442, 1897

\bibitem[{{Ripepi} {et~al}\mbox{.}(2012){Ripepi}, {Moretti}, {Marconi},
  {Clementini}, {Cioni}, {Marquette}, {Girardi}, {Rubele}, {Groenewegen}, {de
  Grijs}, {Gibson}, {Oliveira}, {van Loon}, \& {Emerson}}]{retal12b}
---, 2012, \mnras, 424, 1807

\bibitem[{{Rubele} {et~al}\mbox{.}(2015){Rubele}, {Girardi}, {Kerber}, {Cioni},
  {Piatti}, {Zaggia}, {Bekki}, {Bressan}, {Clementini}, {de Grijs}, {Emerson},
  {Groenewegen}, {Ivanov}, {Marconi}, {Marigo}, {Moretti}, {Ripepi},
  {Subramanian}, {Tatton}, \& {van Loon}}]{retal15}
{Rubele} S. {et~al.}, 2015, ArXiv e-prints

\bibitem[{{Rubele} {et~al}\mbox{.}(2012){Rubele}, {Kerber}, {Girardi}, {Cioni},
  {Marigo}, {Zaggia}, {Bekki}, {de Grijs}, {Emerson}, {Groenewegen},
  {Gullieuszik}, {Ivanov}, {Miszalski}, {Oliveira}, {Tatton}, \& {van
  Loon}}]{retal12}
---, 2012, \aap, 537, A106

\bibitem[{{Sabbi} {et~al}\mbox{.}(2009){Sabbi}, {Gallagher}, {Tosi},
  {Anderson}, {Nota}, {Grebel}, {Cignoni}, {Cole}, {Da Costa}, {Harbeck},
  {Glatt}, \& {Marconi}}]{setal09}
{Sabbi} E. {et~al.}, 2009, \apj, 703, 721

\bibitem[{{Skowron} {et~al}\mbox{.}(2014){Skowron}, {Jacyszyn}, {Udalski},
  {Szyma{\'n}ski}, {Skowron}, {Poleski}, {Koz{\l}owski}, {Kubiak},
  {Pietrzy{\'n}ski}, {Soszy{\'n}ski}, {Mr{\'o}z}, {Pietrukowicz}, {Ulaczyk}, \&
  {Wyrzykowski}}]{setal14}
{Skowron} D.~M. {et~al.}, 2014, ArXiv e-prints

\bibitem[{{Stetson}, {Davis} \& {Crabtree}(1990){Stetson}, {Davis}, \&
  {Crabtree}}]{setal90}
{Stetson} P.~B., {Davis} L.~E., {Crabtree} D.~R., 1990, in Astronomical Society
  of the Pacific Conference Series, Vol.~8, CCDs in astronomy, {Jacoby} G.~H.,
  ed., pp. 289--304

\bibitem[{{Tatton} {et~al}\mbox{.}(2013){Tatton}, {van Loon}, {Cioni},
  {Clementini}, {Emerson}, {Girardi}, {de Grijs}, {Groenewegen}, {Gullieuszik},
  {Ivanov}, {Moretti}, {Ripepi}, \& {Rubele}}]{tetal13}
{Tatton} B.~L. {et~al.}, 2013, \aap, 554, A33

\bibitem[{{Testor} {et~al}\mbox{.}(2010){Testor}, {Lemaire},
  {Heydari-Malayeri}, {Kristensen}, {Diana}, \& {Field}}]{tetal10c}
{Testor} G., {Lemaire} J.~L., {Heydari-Malayeri} M., {Kristensen} L.~E.,
  {Diana} S., {Field} D., 2010, \aap, 510, A95

\bibitem[{{Udalski}(2003)}]{u03}
{Udalski} A., 2003, \actaa, 53, 291

\bibitem[{{Zaritsky} {et~al}\mbox{.}(2002){Zaritsky}, {Harris}, {Thompson},
  {Grebel}, \& {Massey}}]{zetal02}
{Zaritsky} D., {Harris} J., {Thompson} I.~B., {Grebel} E.~K., {Massey} P.,
  2002, \aj, 123, 855

\end{thebibliography}
%
%

\clearpage

\begin{table}
\caption{VMC tile information.}
\begin{tabular}{@{}lccccccccc}\hline
Tile ID & RA (J2000) & Dec (J2000) & Completion in      & \multicolumn{3}{c}{mag$_{\rm 0.1}$} & 
\multicolumn{3}{c}{50\% Completeness level} \\ 
          &  ($\degr$)  & ($\degr$)  & the $K_{\rm s}$ filter  & $Y$  & $J$  & $K_{\rm s}$ & $Y$  & $J$  & $K_s$ \\\hline
SMC 3$\_$5 &  21.9762 & $-$74.0018 & 100\%                & 21.73 & 21.70 & 20.62 & 22.24 & 22.09 &  21.02     \\         
SMC 4$\_$5 &  21.2959 & $-$72.9339 & 100\%                 & 22.09 & 21.72 & 20.65 & 22.43 & 22.15 &  21.08     \\    
SMC 5$\_$6 &  25.4401 & $-$71.5879 & 92\%                 & 22.02 & 21.70 & 20.61 & 22.52 & 22.12 &  20.98     \\        
SMC 6$\_$5 &  20.4138 & $-$70.7601 & 100\%                 & 22.07 & 21.58 & 20.59 & 22.48 & 22.20 &  20.93     \\         
BRI 2$\_$3 &  33.6941 & $-$74.0132 & 92\%                 & 21.79 & 21.62 & 20.68 & 21.93 & 21.71 &  20.96     \\    
\hline
\end{tabular}
\end{table}

\begin{table*}
\caption{Fundamental parameters of the SMC clusters studied.}
\begin{tabular}{@{}lccccccc}\hline
Name & RA       & Dec       & Diameter   & $E(B-V)$ &  log($t$ yr$^{-1}$)  & Z & Comments \\ 
     & ($\degr$)  & ($\degr$)  & ($\arcmin$)  & (mag)  &            &   & \\\hline
\multicolumn{8}{c}{SMC 3$\_$5} \\\hline
GKH51*          &  022.277 & $-$73.5635& 0.30 &  0.08& ... & ...   & possible non-cluster (1)   \\     
B165           &  022.711 & $-$73.434 & 0.45 &  0.08& ... & ...   & possible non-cluster   \\  
B166           &  022.975 & $-$73.910 & 0.60 &  0.08& 8.3 & 0.003 &    \\         
L109           &  023.306 & $-$74.167 & 0.90 &  0.07& 9.6 & 0.003 & 9.55$\pm$0.10 (2),  9.50$\pm$0.05 (7)  \\  
SGDH-cluster A &  022.308 & $-$73.532 & 0.40 &  0.06& ... & ...   & possible non-cluster (1)  \\  
B164           &  022.369 & $-$73.533 & 0.60 &  0.08& 8.5 & 0.003 &    \\  
GKH54/57*       &  022.400 & $-$73.551 & 0.10 &  0.08& ... & ...   & possible non-cluster (1)   \\  
GKH29*          &  022.400 & $-$73.558 & 0.10 &  0.08& ... & ...   & possible non-cluster (1)   \\  
GKH24*          &  022.412 & $-$73.555 & 0.05 &  0.08& ... & ...   & not resolved (1)   \\  
GKH22          &  022.421 & $-$73.563 & 0.40 &  0.08& ... & ...   & not resolved (1)   \\  
HW75           &  019.372 & $-$73.570 & 0.90 & 0.06 & 8.3 & 0.003 & 8.20$\pm$0.40 (6)   \\  
B155           &  020.099 & $-$74.004 & 0.50 & 0.04 & 8.7 & 0.003 &    \\\hline 
\multicolumn{8}{c}{SMC 4$\_$5} \\\hline
HW74           &  019.194 & $-$73.1605& 0.55&$>$0.06& 8.8 & 0.003 & 7.50$\pm$0.20 (6)  \\  
B156           &  019.891 & $-$73.0967& 0.65 &  0.06& ... & ...   & possible non-cluster; 7.50$\pm$0.20 (6)     \\  
HW72           &  018.923 & $-$73.167 & 0.50 &  0.06& 8.7 & 0.003 & 8.10$\pm$0.20 (6)   \\  
K68,L98        &  018.890 & $-$72.624 & 0.70 &0.05  & 8.7 & 0.003 & 8.20$\pm$0.20 (6)    \\  
B162           &  020.963 & $-$73.440 & 0.55 & 0.06 & 7.3 & 0.003 &    \\  
H86-211        &  021.191 & $-$73.425 & 0.50 &0.06  & ... & ...   &  possible non-cluster   \\  
H86-212        &  021.313 & $-$73.501 & 0.60 &0.06  & ... & ...   &  possible non-cluster   \\  
BS282          &  021.375 & $-$73.3895& 0.40 &0.06  & ... & ...   &  possible non-cluster  \\  
HW77           &  020.044 & $-$72.622 & 1.40 &0.06  & 9.15& 0.003 &    \\  
HW78           &  020.335 & $-$73.094 & 0.45 &0.06  & ... & ...   &  possible non-cluster  \\  
HW81           &  021.042 & $-$73.154 & 1.00 &$>$0.3& 7.0& 0.003 &    \\  
BS176          &  021.063 & $-$73.1685& 0.35 &0.06  & 8.3 & 0.003 & \\  
L110           &  023.608 & $-$72.874 & 2.40 &  0.06& 9.8 & 0.003 & 9.80$\pm$0.10 (3), 9.90$\pm$0.10 (7) \\  
H86-213        &  023.673 & $-$73.2755& 0.40 &  0.07& 8.8 & 0.003 &    \\  
HW82           &  021.114 & $-$73.172 & 0.70 &0.06  & 7.8 & 0.003 &    \\  
HW80           &  020.858 & $-$73.224 & 0.70 &0.06  & ... & ...   & possible non-cluster   \\  
BS187          &  022.754 & $-$72.851 & 0.60 &  0.04& 9.3 & 0.003 &    \\  
Sk158          &  018.971 & $-$73.319 & 0.70 &0.06  & 7.3 & 0.003 &    \\  
Sk157          &  018.965 & $-$73.347 & 0.60 &0.06  & 7.2 & 0.003 &    \\\hline
\multicolumn{8}{c}{SMC 5$\_$6} \\\hline
BS188          &  023.796 & $-$71.737 & 0.80 &  0.07& 9.2 & 0.003 &    \\  
HW84           &  025.431 & $-$71.162 & 1.20 &  0.07& 9.25& 0.003 & 9.20$\pm$0.10 (2)   \\  
HW85           &  025.615 & $-$71.2795& 0.70 &  0.07& 9.4 & 0.003 & 9.30$\pm$0.10 (2)   \\  
BS189          &  025.777 & $-$71.452 & 1.20 &  2.80& 7.3 & 0.003 &     \\  
BS190          &  025.963 & $-$71.747 & 1.20 &  0.50& 8.5 & 0.003 &     \\\hline 
\multicolumn{8}{c}{SMC 6$\_$5} \\\hline
HW73           &  019.105 & $-$71.327 & 0.80 &0.05  & 8.3 & 0.003 &  8.15$\pm$0.20 (6)   \\  
L95            &  018.690 & $-$71.348 & 0.50 &0.05  & 8.4 & 0.003 &  8.30$\pm$0.20 (6)   \\  
H86-197        &  018.882 & $-$71.176 & 0.50 &  0.06& 9.3 & 0.006 & 9.10$\pm$0.10 (5)  \\  
HW67           &  018.259 & $-$70.964 & 0.95 &  0.06& 9.2 & 0.003 & 9.20$\pm$0.10 (4)  \\  
BS173          &  021.017 & $-$70.327 & 1.00 &  0.06& ... &  ...  & possible non-cluster    \\  
B168           &  021.678 & $-$70.785 & 0.50 &  0.06& 9.6 & 0.003 &    \\  
IC1708         &  021.233 & $-$71.185 & 1.00 &  0.07& 9.1 & 0.003 & 9.10$\pm$0.10 (2)   \\\hline
\multicolumn{8}{c}{BRI 2$\_$3} \\\hline
BS226          &  031.425 & $-$74.381 & 0.40 &  ... &  ...& ...   & not recognized  \\  
BS229          &  031.952 & $-$74.4425& 0.50 &  0.08&  7.3& 0.003 &    \\  
BS232          &  032.340 & $-$74.026 & 0.50 &  0.08&  ...& ...   & possible non-cluster   \\  
BS233          &  032.658 & $-$74.156 & 1.20 &  0.15&  7.3& 0.003 &    \\  
BS235          &  032.962 & $-$74.119 & 0.60 &  0.09&  ...& ...   &   possible non-cluster  \\  
BS239          &  033.651 & $-$73.985 & 1.20 &  0.11&  8.5& 0.003 &    \\  
BSBD4          &  033.667 & $-$74.358 & 0.60 &  0.08&  9.1& 0.003 &    \\  
BS240          &  033.717 & $-$73.953 & 0.35 &  0.09&  ...& ...   &   possible non-cluster   \\\hline
\end{tabular}
\end{table*}

\setcounter{table}{1}
\begin{table*}
\caption{continued.}
\noindent * In Table 3 of \citet{betal08} the acronym GHK appears instead of GKH (see their table 1).

\noindent References: (1) \citet{getal07}; (2) \citet[$CT_1$
  data]{pietal11}; (3) \citet[$CT_1$ data]{petal07}; (4) \citet[$CT_1$
  data]{p11b}; (5) \citet[$CT_1$ data]{pb12}; (6) \citet{getal10}; (7) \citet{paetal14}.

\end{table*}

\begin{table}
\caption{Fundamental parameters of additional SMC clusters.}
\begin{tabular}{@{}lccccccccc}\hline
Name &   log($t$ yr$^{-1}$)  & [Fe/H] & Reference \\\hline

BS196 & 9.70$\pm$0.05 &  $-$1.7$\pm$0.1 & 1 \\
HW66  & 9.60$\pm$0.10 &  $-$1.3$\pm$0.2 & 2 \\
HW79  & 9.70$\pm$0.10 &  $-$1.3$\pm$0.2 & 2 \\
HW86  & 9.15$\pm$0.10 &  $-$0.65$\pm$0.20 & 2, 6 \\
L100  & 9.30$\pm$0.10 &  $-$0.7$\pm$0.2 & 2 \\
L106  & 9.20$\pm$0.15 &  $-$0.7$\pm$0.2 & 2 \\
L108  & 9.30$\pm$0.10 &  $-$0.9$\pm$0.2 & 2, 6 \\
L111  & 9.15$\pm$0.10 &  $-$0.75$\pm$0.20 & 2, 6 \\
L112  & 9.75$\pm$0.10 &  $-$1.1$\pm$0.2 & 2, 6 \\
L113  & 9.70$\pm$0.10 &  $-$1.17$\pm$0.12 & 3, 4, 6 \\ 
L114  & 8.15$\pm$0.15 &  $-$0.7$\pm$0.2 & 5 \\ 
L115  & 8.05$\pm$0.10 &  $-$0.7$\pm$0.2 & 5 \\\hline

\end{tabular}
\medskip

\noindent References: (1) \citet{betal08b}; (2) \citet{pietal11}; (3)
\citet{petal07}; (4) \citet{dh98}; (5) \citet{petal07b}; (6) \citet{paetal14}.

\end{table}

\clearpage

\begin{figure*}
\includegraphics[width=144mm]{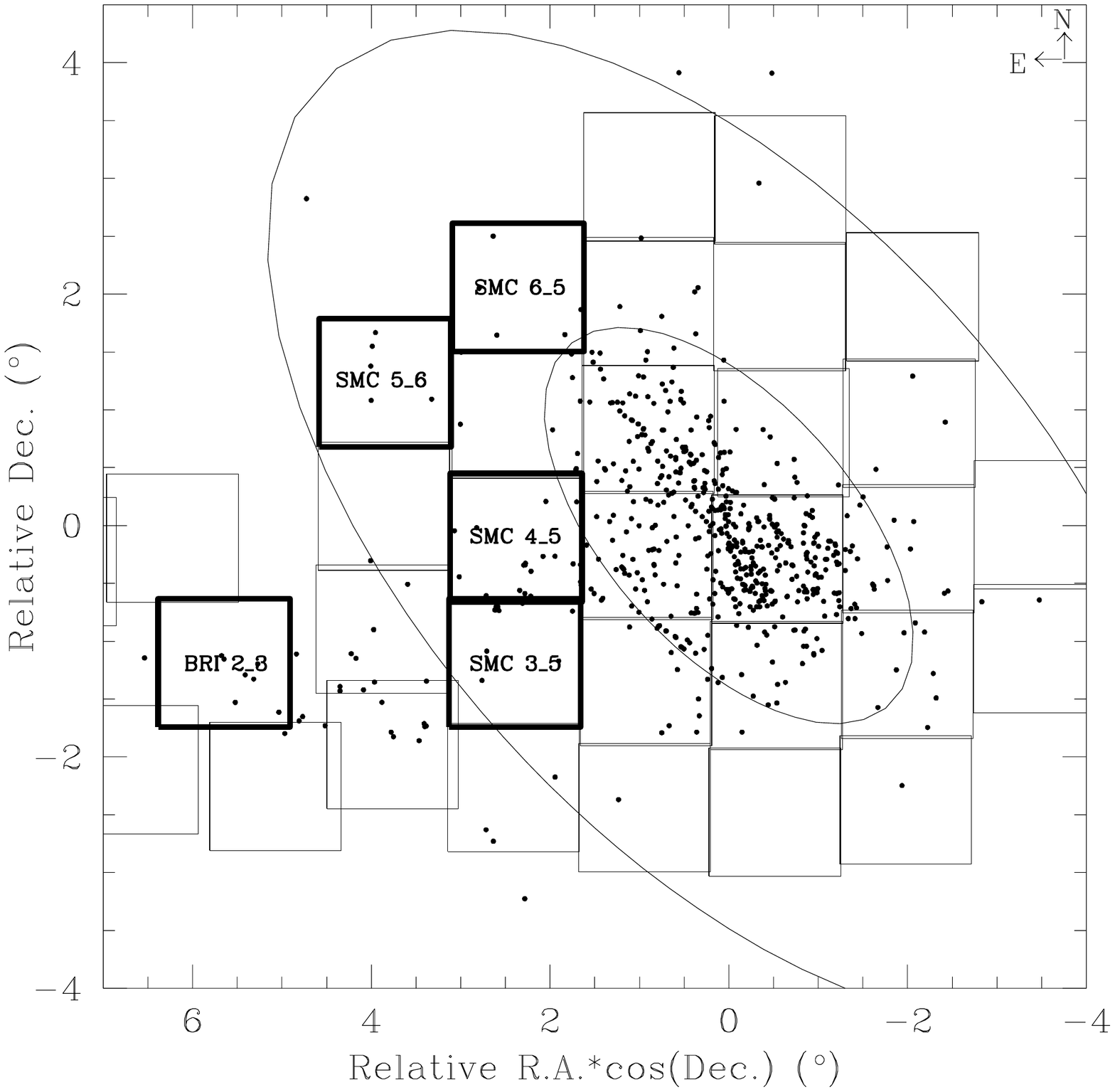}
\caption{Spatial distribution of VISTA tiles across the SMC and the
  western Bridge regions. The distribution of SMC clusters (dots) as
  well as ellipses with semimajor axes of 2.4$\degr$ and 6$\degr$ are
  overplotted. The SMC tiles studied here have been labelled and
  highlighted with thick lines.}
\label{fig1}
\end{figure*}

\clearpage
\begin{figure*}
\includegraphics[width=144mm]{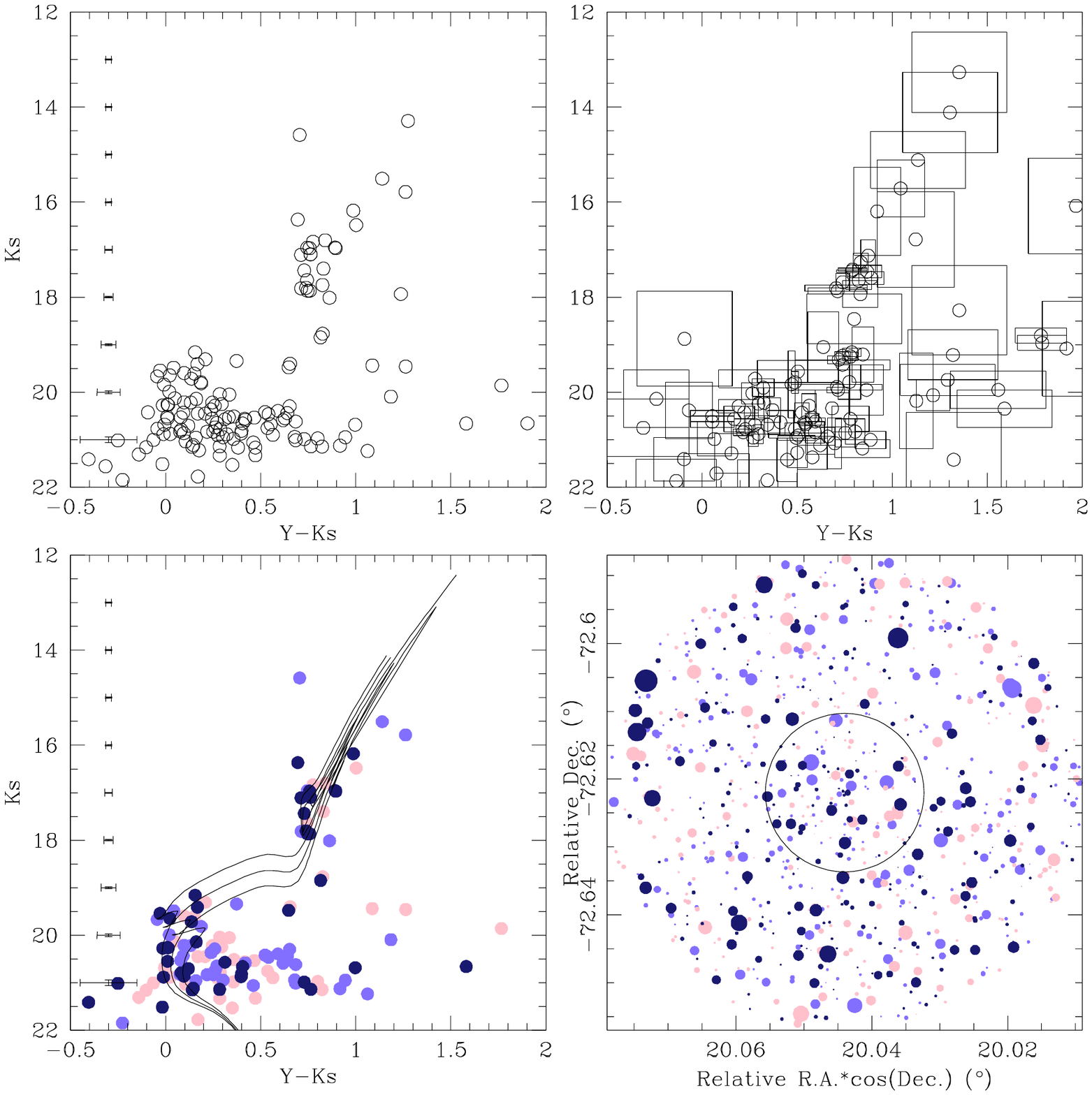}
\caption{CMDs for stars in the field of HW\,77 of tile SMC 4$\_$5: the
  observed CMD composed of the stars distributed within the cluster
  radius (top left-hand panel); a field CMD for an annulus 
 -outer and inner radii equal to 3.163 and 3.0 times the cluster radius-
centred on  the cluster (top right-hand
  panel); the cleaned cluster CMD (bottom left). We overplotted boxes
  for each star in the field CMD to be used in the cluster CMD field
  decontamination (see Section 4 for details). Colour-scaled symbols
  represent stars that statistically belong to the field ($P \le$
  25\%, pink), stars that might belong to either the field or the
  cluster ($P =$ 50\%, light blue), and stars that predominantly
  populate the cluster region ($P \ge$ 75\%, dark blue). Three
  isochrones from \citet{betal12} for log($t$ yr$^{-1}$), log($t$
  yr$^{-1}$) $\pm$ 0.1, and the metallicity values listed in Table 2
  are also superimposed. The schematic diagram centred on the cluster
  for a circle of radius three times the cluster radius is shown in
  the bottom right-hand panel. The black circle represents the adopted
  cluster radius. Symbols are as in the bottom left-hand panel, with
  sizes proportional to the stellar brightnesses. North is up; east is
  to the left.}
\label{fig2}
\end{figure*}

\clearpage

\begin{figure}
\includegraphics[width=144mm]{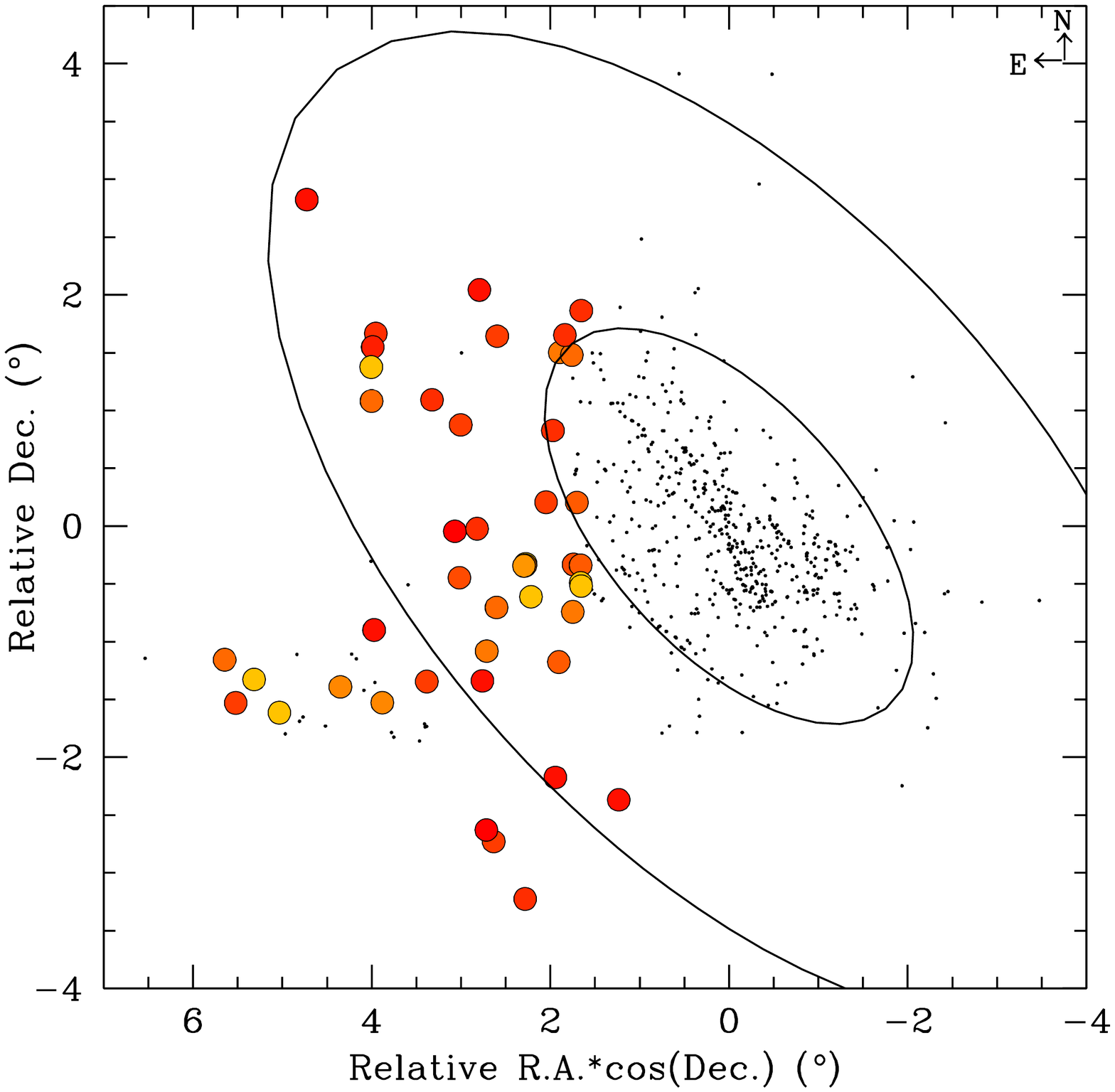}
\caption{Spatial distribution of the enlarged SMC cluster sample
  (solid circles), where darker solid circles correspond to older
  cluster ages. Catalogued clusters (dots) and ellipses with semimajor
  axes of 2.4$\degr$ and 6$\degr$ are overplotted.}
\label{fig3}
\end{figure}

\clearpage

\begin{figure*}
\includegraphics[width=144mm]{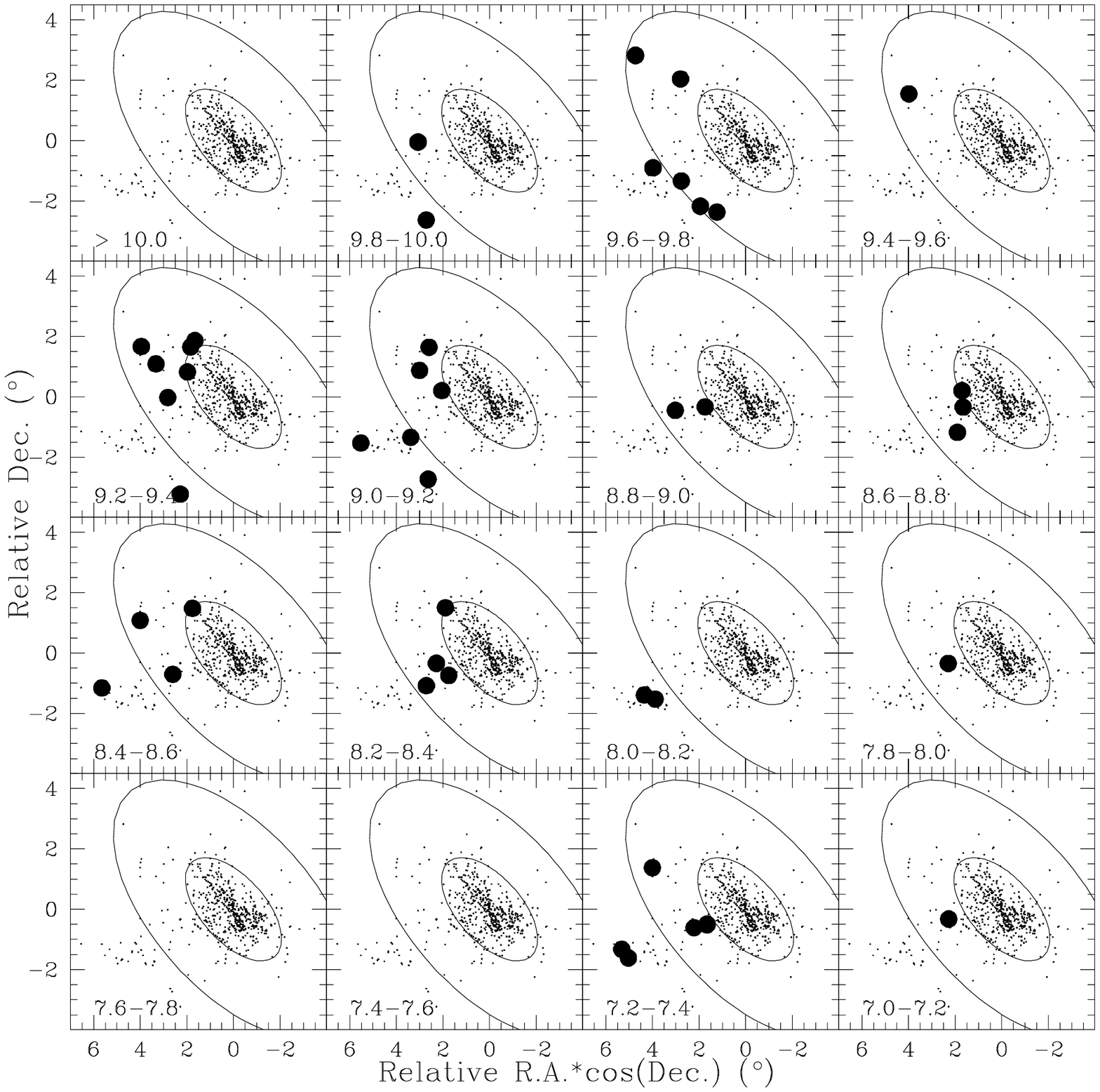}
\caption{As Fig. 3, but for different log($t$ yr$^{-1}$) intervals, as
  labelled in each panel; 
the lower  age limit is included in the respective
interval.}
\label{fig4}
\end{figure*}

\clearpage

\begin{figure*}
\includegraphics[width=144mm]{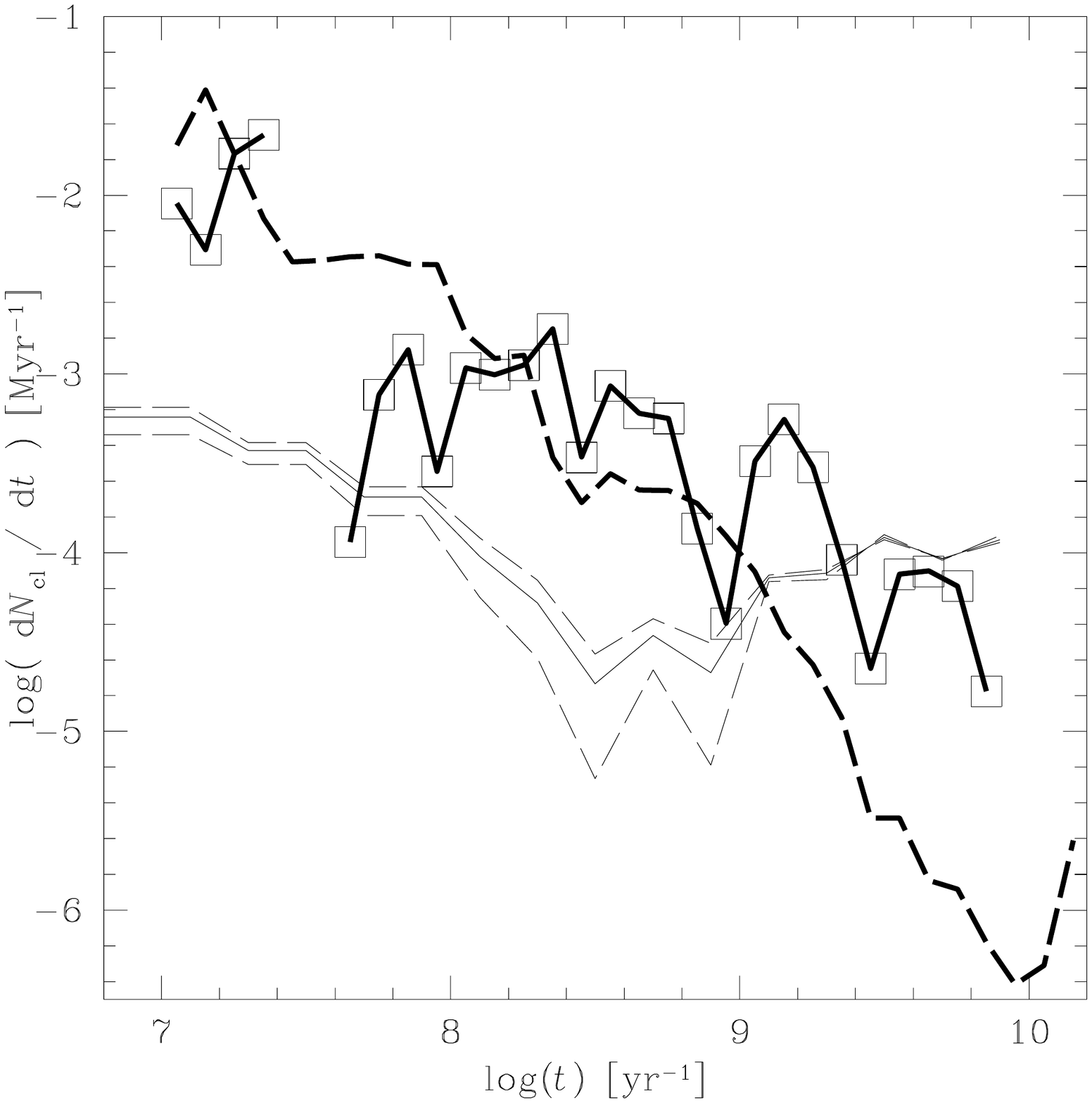}
\caption{Comparison of the observed CFs (open boxes connected by a
thick solid line) and the
  theoretical CF (thin solid line). The error curves for the latter
  are shown as thin dashed lines. The CF constructed on the basis of the 
cluster age determinations of de Grijs \& Goodwin (2008) is overplotted
with a thick dashed line.
 }
\label{fig5}
\end{figure*}



\label{lastpage}
\end{document}